\documentclass[journal=jctcce,manuscript=article, etalmode=truncate]{achemso}
\setkeys{acs}{etalmode=truncate,maxauthors=0,articletitle=true}
\usepackage[version=3]{mhchem} % Formula subscripts using \ce{}
\usepackage[T1]{fontenc}

\usepackage[caption=false]{subfig}
\usepackage{graphicx,color}
\usepackage{amssymb,amsmath,amsthm,amsfonts,amsbsy,mathrsfs}
\usepackage{epsfig}
\usepackage[colorlinks   = true, urlcolor     = cyan, linkcolor    = blue, citecolor   = magenta]{hyperref}
\usepackage{algorithm}
\usepackage{epstopdf}

\usepackage{tikz}
\usepackage{pgfplots}
\usepgfplotslibrary{external}
\usetikzlibrary{shapes.geometric, arrows}

\pgfplotsset{
    compat=newest,
    table/header=false,
    title style={font=\small},
    tick label style={font=\scriptsize},
    label style={font=\scriptsize},
    legend style={font=\scriptsize},
    legend cell align=left
}

\usepackage{tabularx,multirow,booktabs}
\newcolumntype{L}{>{\raggedright\arraybackslash}X}
\newcolumntype{C}{>{\centering\arraybackslash}X}
\newcolumntype{R}{>{\raggedleft\arraybackslash}X}

\tikzstyle{startstop} = [rectangle, rounded corners, line width = 0.6pt, minimum width=3cm, minimum height=1cm,text centered, draw=black, fill=red!30]
\tikzstyle{long_startstop} = [rectangle, rounded corners, line width = 0.6pt, minimum width=4cm, minimum height=1cm, text width=4cm, text centered, draw=black, fill=red!30]
\tikzstyle{io} = [trapezium, trapezium left angle=70, trapezium right angle=110, line width = 0.6pt, minimum width=3cm, minimum height=1cm, text centered, draw=black, fill=blue!30]
\tikzstyle{process} = [rectangle, line width = 0.6pt, minimum width=3cm,  text width=3cm, minimum height=1cm, text centered, draw=black, fill=orange!30]
\tikzstyle{short_process} = [rectangle, line width = 0.6pt, minimum width=2.5cm,  text width=2.5cm, minimum height=1cm, text centered, draw=black, fill=orange!30]
\tikzstyle{long_process} = [rectangle, minimum width=4cm, line width = 0.6pt, minimum height=1cm, text centered, text width=4cm, draw=black, fill=orange!30]
\tikzstyle{very_long_process} = [rectangle, minimum width=5cm, line width = 0.6pt, minimum height=1cm, text centered, text width=5cm, draw=black, fill=orange!30]
\tikzstyle{decision} = [diamond, line width = 0.6pt, minimum width=3cm, minimum height=1cm, text centered, aspect=3, draw=black, fill=green!30]
\tikzstyle{arrow} = [thick,->,>=stealth]

\newcommand{\beq}{\begin{equation}}
\newcommand{\eeq}{\end{equation}}
\newcommand{\beqs}{\begin{eqnarray}}
\newcommand{\eeqs}{\end{eqnarray}}
\newcommand{\beql}{\begin{equation} \label}

\newcommand{\calI}{{\cal I}}

\newcommand{\norm}[2]{\lVert#1\rVert_{#2}}

\let\oldFootnote\footnote
\newcommand\nextToken\relax

\renewcommand\footnote[1]{%
    \oldFootnote{#1}\futurelet\nextToken\isFootnote}

\newcommand\isFootnote{%
    \ifx\footnote\nextToken\textsuperscript{,}\fi}

\author{Amartya S. Banerjee}
\email{asb@lbl.gov} 
\affiliation[LBL]{Computational Research Division,
Lawrence Berkeley National Laboratory, Berkeley, California 94720, U.S.A}

\author{Lin Lin}
\email{linlin@math.berkeley.edu} 
\affiliation{Department of
Mathematics, University of California, Berkeley, California 94720, U.S.A}
\alsoaffiliation{Computational Research Division,
Lawrence Berkeley National Laboratory, Berkeley, California 94720, U.S.A}

\author{Phanish Suryanarayana}
\email{phanish.suryanarayana@ce.gatech.edu}
\affiliation{College of Engineering, Georgia Institute of Technology, Atlanta, Georgia 30332, U.S.A}

\author{Chao Yang}
\email{cyang@lbl.gov} 
\affiliation[LBL]{Computational Research Division, Lawrence Berkeley National Laboratory, Berkeley, California 94720, U.S.A}

\author{John E. Pask}
\email{pask1@llnl.gov}
\affiliation{Physics Division, Lawrence Livermore National Laboratory, Livermore, California 94550, U.S.A}

\title[Complementary_Subspace]{Two-level Chebyshev filter based complementary subspace method: pushing the envelope of large-scale electronic structure calculations}

\begin{document}
\begin{figure}
\subfloat{\scalebox{0.8}
{\begin{tikzpicture} 
\pgfkeys{
    /pgf/number format/precision=1, 
    /pgf/number format/fixed zerofill=true
}
\begin{axis}[ybar stacked,font=\sffamily, 
width=\textwidth,
height=0.5\textwidth,
%xmajorticks=false,
bar width=20pt,
ymax = 210,
nodes near coords,
%every text node part/.style={align=center},
every node near coord/.style={
      check for zero/.code={
        \pgfkeys{/pgf/fpu=true}
        \pgfmathparse{\pgfplotspointmeta-4.5}
        \pgfmathfloatifflags{\pgfmathresult}{-}{
           \pgfkeys{/tikz/coordinate}
        }{}
        \pgfkeys{/pgf/fpu=false}
      }, check for zero, font=\tiny},
%every node near coord/.append style={font=\tiny},
y label style={yshift=-5pt},
ylabel={Wall time (s)},
xtick=data,
xticklabels={Std.~CheFSI, CS2CF,  Std.~CheFSI, CS2CF, Std.~CheFSI, CS2CF, Std.~CheFSI, CS2CF, Std.~CheFSI, CS2CF, Std.~CheFSI, CS2CF},
label style={font=\sffamily\scriptsize},
tick label style={font=\sffamily\scriptsize},
x tick label style={rotate=50,anchor=east, yshift= -5.5, xshift = 2.5},
legend style={at={(0.5,-0.25)},  anchor=north,legend columns=1,
fill=none,font=\sffamily, cells={anchor=west},row sep=1pt}] 
% Plot data
\addplot [fill=brown!35,] coordinates {(10, 12.3) (50,0) (110, 18.3) (150, 7.6) (210, 22.9) (250, 10.8) (310, 21.3) (350, 4.4) (410, 33.0) (450,14.8) (510, 47.4) (550, 25.9)}; 
\addplot [fill=black!25,] coordinates  {(10, 20.7) (50,0.5) (110, 21.2) (150,1.2) (210, 26.8) (250, 2.1) (310, 25.6) (350, 1.6) (410, 30.9) (450,2.4) (510, 125.5) (550,4.2)};

% % Print the totals
\draw (10,50) node [font=\sffamily, draw = none, align=left,   below] {\tiny{33.0}};
\draw (50, 17.5) node [font=\sffamily, draw = none, align=left,   below] {\tiny{0.5}};

\draw (110,57) node [font=\sffamily, draw = none, align=left,   below] {\tiny{39.5}};
\draw (150, 25.5) node [font=\sffamily, draw = none, align=left,   below] {\tiny{8.8}};

\draw (210,66.7) node [font=\sffamily, draw = none, align=left,   below] {\tiny{49.7}};
\draw (250, 29.9) node [font=\sffamily, draw = none, align=left,   below] {\tiny{12.9}};

\draw (310, 63.9) node [font=\sffamily, draw = none, align=left,   below] {\tiny{46.9}};
\draw (350, 23.0) node [font=\sffamily, draw = none, align=left,   below] {\tiny{6.0}};

\draw (410, 80.9) node [font=\sffamily, draw = none, align=left,   below] {\tiny{63.9}};
\draw (450, 34.2) node [font=\sffamily, draw = none, align=left,   below] {\tiny{17.2}};

\draw (510,189.9) node [font=\sffamily, draw = none, align=left,   below] {\tiny{172.9}};
\draw (550, 47.1) node [font=\sffamily, draw = none, align=left,   below] {\tiny{30.1}};

%% Print some labels
\draw (45 ,65.0) node [font=\sffamily, align=left,   below] {\tiny{\begin{tabular}{c} {Electrolyte3D}\\ system \end{tabular}}};

\draw (150 ,70.0) node [font=\sffamily, align=left,   below] {\tiny{\begin{tabular}{c} {SiDiamond3D}\\ system \end{tabular}}};

\draw (250 , 75.0) node [font=\sffamily, align=left,   below] {\tiny{\begin{tabular}{c} {Graphene2D}\\ system \end{tabular}}};

\draw (350 + 3, 85.0) node [font=\sffamily, align=left,   below] {\tiny{\begin{tabular}{c} {Graphene2D}\\ system with \\ fewer extra \\ and  top \\ states\end{tabular}}};

\draw (450 , 90.0) node [font=\sffamily, align=left,   below] {\tiny{\begin{tabular}{c} {CuFCC3D}\\ system \end{tabular}}};

\draw (550 , 200.0) node [font=\sffamily, align=left,   below] {\tiny{\begin{tabular}{c} {LiBCC3D}\\ system \end{tabular}}};

% % Legend
\legend{\tiny{\begin{tabular}{l}Subspace diagonalization (Standard CheFSI) or\\ computation of top states (CS2CF strategy)\end{tabular}}, \tiny{\begin{tabular}{l}Subspace rotation and misc.~calculations (Standard CheFSI) or\\ misc.~calculations (CS2CF strategy) \end{tabular}}}

% % Insert a table here
\draw (200 , 187.0) node [font=\sffamily, align=left,   below] {\scriptsize{\begin{tabular}{| l |}
\hline \\
{Electrolyte3D system:} {$8,586$ atoms, $29,808$ electrons, $1,728$ MPI processes}\\
{SiDiamond3D system:} {$8,000$ atoms, $32,000$ electrons, $1,728$ MPI processes}\\
{Graphene2D system:} {$11,520$ atoms, $23,040$ electrons, $2,304$ MPI processes}\\
{CuFCC3D system:} {$4,000$ atoms, $44,000$ electrons, $1,000$ MPI processes}\\
{LiBCC3D system:} {$27,648$ atoms, $82,944$ electrons, $6,480$ MPI processes} \\
\\ \hline
\end{tabular}}};
\end{axis} 
 \end{tikzpicture}}
}
 \caption*{\footnotesize{(Graphic for the Table of Contents / Graphical Abstract) %Comparison of the wall times of a few key steps in the standard CheFSI (Rayleigh-Ritz and subspace rotation steps) and the CS2CF (computation of top states and other miscellaneous calculations) strategies, for a few large systems. 
 }}
\label{fig:Block_Chart_1_Abstract}
\end{figure}
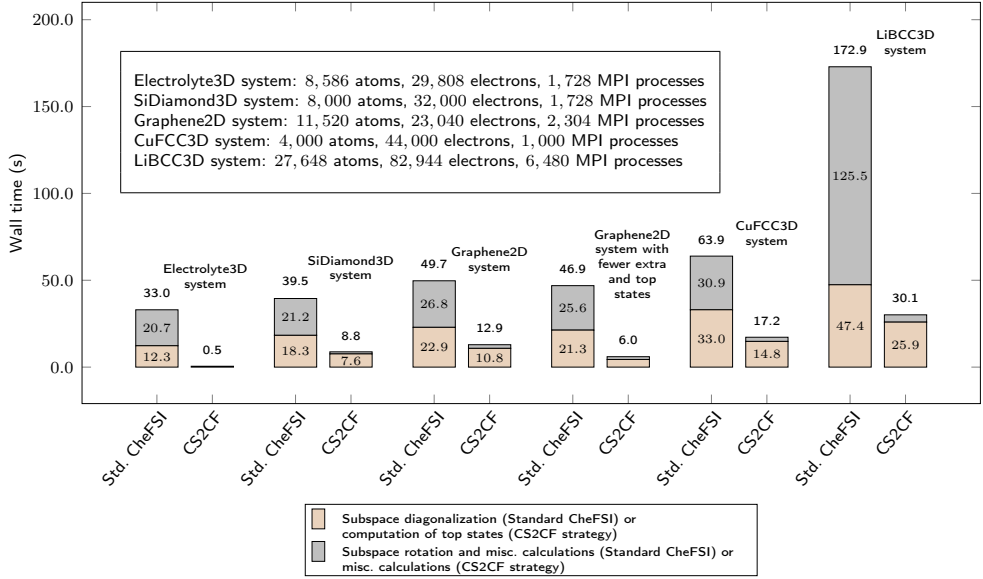
\newpage
\begin{abstract}
We describe a novel iterative strategy 
for Kohn-Sham density functional theory calculations 
aimed at large systems ($> 1000$ electrons), 
applicable to metals and insulators alike.
In lieu of explicit diagonalization of the
Kohn-Sham Hamiltonian on every self-consistent field (SCF) iteration, 
we employ a two-level Chebyshev polynomial filter based complementary subspace strategy to: 
1) compute a set of vectors that span the occupied subspace of the Hamiltonian; 
2) reduce subspace diagonalization to just partially occupied states; and
3) obtain those states in an efficient, scalable manner via an inner Chebyshev-filter iteration.
By reducing the necessary computation to just partially occupied states, 
and obtaining these through an inner Chebyshev iteration, 
our approach reduces the cost of large metallic calculations significantly, 
while eliminating subspace diagonalization for insulating systems altogether. 
We describe the implementation of the method within the framework of the
Discontinuous Galerkin (DG) electronic structure method and show that
this results in a computational scheme that can effectively tackle bulk
and nano systems containing tens of thousands of electrons, 
with chemical accuracy, 
within a few minutes or less of wall clock time per SCF iteration on large-scale computing platforms. 
We anticipate that our method will be instrumental 
in pushing the envelope of large-scale \textit{ab initio} molecular dynamics. 
As a demonstration of this, we simulate a bulk silicon system containing $8,000$ atoms at finite temperature, and obtain an average SCF step wall time of $51$ seconds on $34,560$ processors; thus allowing us to carry out $1.0$ ps of \textit{ab initio} molecular dynamics in approximately $28$ hours (of wall time).
\end{abstract}
\section{Introduction} \label{sec:Introduction}
Kohn-Sham density functional theory (KS-DFT) \citep{KohnSham_DFT, Martin_ES} is among the most widely used  approaches in the computational chemistry, condensed matter, and materials research communities. Over the years, KS-DFT has provided unparalleled insights and robust predictions for the gamut of materials properties \citep{burke2017density, burke2012perspective, becke2014perspective}, as a result of which, much research has been devoted to enable calculations of ever larger and more complex systems \citep{fattebert2016modeling, bowler2010calculations, vandevondele2012linear,cohen2008insights,  luo2014density}.

Conventionally, the Kohn-Sham equations are solved self-consistently, 
wherein the linear eigenvalue problem arising from discretization in a chosen basis 
is solved on each self-consistent field (SCF) iteration until a fixed point is reached in the electronic density or potential; whereupon energies, forces, and other quantities of interest are computed \cite{Martin_ES}. 
Solution of this eigenvalue
problem via direct or iterative diagonalization methods scales cubically
with the number of electronic states in the system (and hence, also
cubically with the number of atoms). The computational cost of this
procedure can become prohibitive, however, as the system size grows beyond a
few thousand electronic states. This has led to the development of
computational techniques which scale more favorably with respect to the
number of electronic states in the system, by avoiding explicit
diagonalization of the Kohn-Sham Hamiltonian. \citep{goedecker1999linear,
bowler2012methods, JCP_122_084119_2005_ONETEP,
CMS2009,JPCM_25_295501_2013_PEXSI,suryanarayana2017sqdft, pratapa2016spectral,aarons2016perspective} 
However, these work best on insulating systems with substantial
band gaps \citep{aarons2016perspective, ratcliff2017challenges}, metallic
systems with low dimension (such as in the case of the pole
expansion and selected inversion (PEXSI) method) \citep{hu2015dgdft, lin2014siesta},
or systems at high electronic temperature \cite{pratapa2016spectral,suryanarayana2017sqdft}. 
Consequently, while impressive large-scale Kohn-Sham calculations of various
insulating systems have been demonstrated by several groups
\citep{hine2009linear, fattebert2016modeling, bowler2010calculations,
vandevondele2012linear}, large-scale, well converged, chemically
accurate calculations of realistic semiconducting or metallic systems
have appeared only rarely 
\citep{aarons2016perspective, ratcliff2017challenges,
suryanarayana2017sqdft, ruiz2013variational}. Many of the aforementioned
methods rely on the nearsightedness principle
\citep{prodan2005nearsightedness} for obtaining the electron density
from the Kohn-Sham Hamiltonian in an efficient manner. This presents
practical issues while dealing with bulk metallic systems at moderate
electronic temperatures (up to a few thousand Kelvins). The relatively
slow decay of the density matrix associated with such systems (even for
the case of simple metals like aluminum) often results in a
computational method having a favorable algorithmic scaling but 
relatively large prefactor. \citep{suryanarayana2017nearsightedness} 
In practical treatments of large bulk metallic systems containing several thousands of atoms, this is likely to result in a larger computational wall time per SCF iteration with the use of these methods\footnote{Aggressive truncation of the density matrix entries can also lead to SCF iteration instabilities, further slowing calculations with these methods \citep{suryanarayana2017nearsightedness, suryanarayana2017sqdft}.} than with the use of diagonalization based methods\citep{suryanarayana2017nearsightedness}. 

In this work, we take a different approach
to make Kohn-Sham calculations of large, metallic systems more
computationally feasible. 
Our approach is to revert to the use of
(cubically scaling) partial diagonalization of the Kohn-Sham
Hamiltonian via iterative methods, but to ensure that the resulting
computational strategy has a low prefactor (in the sense of
computational complexity estimates) 
and good parallel scaling efficiency. 
In practice, with the use of a sufficiently large number of processors, 
this then enables larger systems sizes and/or shorter simulation wall times 
than attainable heretofore.

With the aforementioned goal in mind, we have focused on two principal strategies for achieving it. The first of these has been to ensure that we use a discretization scheme that is highly efficient. Specifically, we utilize basis functions which can produce systematically improvable, high quality numerical results while keeping the number of basis functions per atom required for doing so, small. To this end, we employ so called Adaptive Local Basis (ALB) functions \citep{lin2012adaptive} that are generated on the fly on every SCF step during the course of Kohn-Sham calculations. These basis functions are able to capture the local materials physics in electronic structure calculations, and when used in conjunction with the interior penalty Discontinuous Galerkin (DG) formalism for the Kohn-Sham equations, they allow high quality energies and Hellman-Feynman forces to be computed with only a few tens of basis functions per atom \citep{lin2012adaptive, zhang2017adaptive,banerjee2016chebyshev}. The DG approach for  solving the Kohn-Sham equations using ALB functions has been incorporated into a massively parallel software package, DGDFT \citep{hu2015dgdft, hu2015edge}. The second strategy has been to employ a well suited iterative diagonalization strategy for the discretized Kohn-Sham Hamiltonian. To this end, we have made use of Chebyshev polynomial filtered subspace iteration (CheFSI) \citep{banerjee2016chebyshev}. For a number of reasons \citep{banerjee2016chebyshev}, this technique works particularly well within the framework of DGDFT --- both in terms of computational wall times and overall parallel scaling efficiency. We have recently demonstrated that the CheFSI technique significantly outperforms existing alternatives in carrying out SCF iterations in DGDFT, particularly in the context of large bulk systems\citep{banerjee2016chebyshev}.

Our experiences with the DGDFT-CheFSI strategy have revealed that the
principal computational bottleneck in treating large, metallic systems 
occurs in the Rayleigh-Ritz process \citep{Saad_large_eigenvalue_book} 
to obtain approximate eigenvalues and vectors in each SCF iteration.
In the context of the DGDFT-CheFSI methodology, this process is generally as follows \citep{banerjee2016chebyshev}: 
\begin{enumerate}\itemsep0pt
\item Compute a basis spanning the occupied subspace using a Chebyshev polynomial filter.
\item Orthonormalize the basis.
\item Project the DG Hamiltonian matrix onto the subspace.
\item Diagonalize the projected Hamiltonian to obtain Ritz values and projected vectors.
\item Rotate the basis according to projected vectors to obtain Ritz vectors.
\end{enumerate}
The Ritz values and vectors so obtained then provide approximate eigenvalues and vectors of the DG Hamiltonian in each SCF iteration of the Kohn-Sham solution. 
In the following, we focus mainly on steps (4) and (5) of the process, which we shall refer to as \emph{subspace diagonalization} and \emph{subspace rotation}, respectively. 
In the absence of fractionally occupied states (i.e., for insulating
systems), these steps can be avoided altogether (see Section
\ref{subsubsec:DM_and_Proj_DM}) and in that case, the computational
bottleneck arises from the orthonormalization of the basis that
spans the occupied subspace. Indeed, due to
their focus on insulating systems, a number of previous authors \citep{car1985unified, bowler2012methods, lam2005computational,
mohr2015accurate} have focused on reducing or eliminating this
orthonormalization cost, instead of the cost associated with the
aforementioned projected Hamiltonian eigenvalue problem. 
For non-insulating (e.g., metallic and semiconducting) systems, however, 
fractional occupation numbers have to be 
computed and the projected Hamiltonian eigenvalue problem cannot be
directly avoided.\footnote{{In principle, it is possible to employ techniques such as Fermi Operator Expansion\citep{goedecker1999linear} (FOE) to arrive at the projected density matrix, starting from the projected Hamiltonian, without going through the intermediate step of computing the eigenstates of the projected Hamiltonian \citep{motamarri2014subquadratic}. 
However, there are a number of challenges in taking such an approach. 
The projected Hamiltonian is in general a dense matrix, especially for metallic systems at moderate electronic temperatures\citep{suryanarayana2017nearsightedness} (e.g., less than $3000$ Kelvin). Furthermore, the FOE technique requires a large number of matrix-matrix products to be computed\citep{motamarri2014subquadratic}. Moreover, the method becomes less efficient at lower (e.g., ambient) electronic temperatures due to the increased number of terms in the expansion. Finally, an efficient application of the FOE technique requires a localization procedure to be carried out, which can be computationally expensive\citep{motamarri2014subquadratic} and can also lead to SCF convergence issues \citep{Carlos_linear}. Due to these factors, application of the FOE technique is expected to be significantly more expensive than the methodology developed in this work (based on computing only the topmost states of the projected Hamiltonian using iterative methods, and so involving only matrix-vector products). Indeed, for these reasons, it is expected that such application of the FOE technique would be more expensive than direct computation of the eigenstates of the projected Hamiltonian, and so we do not pursue this direction here.}} In this situation, detailed wall time studies (e.g., Section \ref{subsec:efficiency_and_scaling}) reveal
that the most significant hindrance to pushing the computational
efficiency of the DGDFT-CheFSI approach lies in the {subspace diagonalization} 
and {subspace rotation} steps of the Rayleigh-Ritz process. While the orthonormalization cost can
become significant for large problems, its contribution to the
simulation wall time is less than that of the above steps, and its parallel scalability is appreciably better. This provides the incentive to devise a computational strategy that reduces or eliminates the computational cost incurred due to the {subspace diagonalization} and {subspace rotation} steps in the DGDFT-CheFSI approach.

In this work, we formulate and implement a two-level Chebyshev polynomial filter based
complementary subspace strategy to address the above issues. In this methodology, only the relatively few fractionally occupied states of the projected Hamiltonian, and not those fully occupied or empty, are calculated, thus reducing the computational cost of the {subspace diagonalization} and {subspace rotation} steps significantly. Moreover, exploiting the spectral properties of the projected Hamiltonian, we employ  CheFSI iterations to obtain the fractionally occupied states, thereby yielding a highly efficient iterative scheme that uses Chebyshev polynomial filtering on two levels. We refer to the resulting computational methodology as CS2CF, i.e., \emph{Complementary Subspace strategy with 2 levels of Chebyshev Filtering}.

The idea of exploiting completeness to reduce subspace computations to just fractionally occupied states was presented recently by Michaud-Rioux et al.\citep{michaud2016rescu} in their partial Rayleigh-Ritz method for large-scale Kohn-Sham calculations, as implemented in the RESCU Matlab electronic structure code.\footnote{There are significant differences, however, between the partial Rayleigh-Ritz method of Michaud-Rioux et al.\citep{michaud2016rescu} and the CS2CF approach presented here. Specifically, in their implementation of the partial Rayleigh-Ritz method, the authors have employed direct diagonalization and/or the iterative LOBPCG method \citep{LOBPCG_1} to obtain the fractionally occupied states, neither of which appears to yield significant savings in the large-scale parallel context. Furthermore, their formulation, as described, is limited to the calculation of energies alone and implementation is limited to MATLAB \citep{MATLAB}. In contrast, the use of an inner-level of CheFSI for computing the fractionally occupied states turns out to be particularly well suited to large-scale parallel implementation for a number of reasons (see Section \ref{subsubsec:chefsi_top_states}) and increases the computational efficiency of the procedure dramatically. Additionally, the CS2CF strategy integrates particularly well within the framework of the massively parallel DGDFT code (see  Section \ref{subsec:DGDFT_Implementation}). Consequently, it allows us to attack metallic systems containing tens of thousands of electrons in a few minutes or lesser of wall time per SCF step (See Section \ref{sec:Results}). Finally, all the results reported here are done with full chemical accuracy (i.e., the discretization parameters were chosen to result in well converged energies and forces), which enables us to carry out \textit{ab initio} molecular dynamics simulations of large scale non-insulating systems conveniently.
}  
The idea has also been exploited previously in particle-hole duality formulations to accelerate density matrix computations in $O(N)$ electronic structure methods \citep{mazziotti2003towards,mazziotti2002variational,truflandier2016communication}. 
In the context of subspace diagonalization, in line with the key idea of obtaining a large subspace from its much smaller complement, we refer to the approach as the \emph{Complementary Subspace} (CS) strategy here.

Once the subspace computation has been reduced to just fractionally occupied states, it is then crucial to obtain these as efficiently as possible. As we detail in Section \ref{subsubsec:chefsi_top_states}, to accomplish this, we exploit key characteristics of the subspace eigenvalue problem to obtain the fractionally occupied states both efficiently and scalably via low-order CheFSI iterations. This strategy is particularly well suited to large-scale parallel implementation and, as we show, 
integrates particularly well within the framework of the massively parallel DGDFT code. 
Consequently, it allows us to attack metallic systems containing tens of thousands of electrons in a few minutes or less of wall time per SCF iteration. Finally, all results reported here are computed to chemical accuracy 
(i.e., energy and force errors below $10^{-3}$ Ha/atom and $10^{-3}$ Ha/Bohr, respectively), 
as typical in production simulations. 
This has allowed us to carry out accurate, energy-conserving 
\textit{ab initio} molecular dynamics simulations of 
systems with many thousands of atoms in a few minutes per MD step. 
Without the use of the CS2CF strategy, the advantages afforded by the ALB discretization, and the highly efficient, massively parallel DGDFT code, this would not be possible.

It is important to note that the utility of the complementary subspace strategy (and in particular, the CS2CF methodology) is not restricted to the DGDFT code. \emph{Any} other Kohn-Sham code can benefit from this methodology as well, although as mentioned above (also see Section \ref{subsec:DGDFT_Implementation}) the CS2CF strategy integrates well and performs particularly efficiently within DGDFT. Overall, we view this work as the development of a Kohn-Sham equation solution strategy that has been well tuned to minimize wall times in practical simulation scenarios of large systems. In our view, it is an important step towards a robust and efficient methodology for carrying out \textit{ab initio} molecular dynamics of large metallic and semiconducting systems.

The remainder of this paper is organized as follows. In Section \ref{sec:Methodology}, we outline the the mathematical formulation of the complementary subspace strategy, efficient two-level Chebyshev filter based solution, and large-scale parallel implementation in the context of the discontinuous Galerkin electronic structure method. In Section \ref{sec:Results}, we present results for a range of systems, from insulating to semiconducting to metallic, and comparisons with existing methods. We conclude and comment on future research directions in Section \ref{sec:Conclusions}. 
\section{Methodology}
\label{sec:Methodology}
We describe the formulation of the complementary subspace strategy as
well as its implementation within the adaptive local basis set based
discontinuous Galerkin electronic structure method (specifically, the
DGDFT code \citep{hu2015dgdft}) in this section. For 
simplicity, we consider $\Gamma$-point calculations of non-spin-polarized 
periodic systems, as is typical in large-scale \textit{ab
initio} molecular dynamics simulations, although this assumption is not required in what follows. 

\subsection{Formulation of complementary subspace strategy}
\label{subsec:formulation}

\subsubsection{Density matrix and projected density matrix}
\label{subsubsec:DM_and_Proj_DM}
We consider a system with a discretized Kohn-Sham Hamiltonian matrix $H$
of size $N_b \times N_b$, with $N_b$ denoting the total number of
(orthonormal) basis functions used for the discretization. The Hamiltonian is a function of the real-space electron density $\rho(x)$ from the discretized Hamiltonian that must be determined iteratively in a self-consistent field (SCF) iteration. 

The conventional procedure for achieving this is through the intermediate computation of the (discretized) Kohn-Sham orbitals, i.e., the eigenvectors of $H$. \citep{Hutter_abinitio_MD, Martin_ES} Within the above setting, each Kohn-Sham orbital is a $N_b \times 1$ sized real valued vector. For a system containing $N_e$ electrons (per unit cell), the lowest $N_s$ eigenstates of $H$ need to be computed in each SCF step. For an insulator, each orbital is doubly occupied, and $N_s$ can be taken as  $N_e / 2$. In contrast, for a metallic system, it is customary to use $N_s = N_e / 2 + N_{x}$, where $N_{x}$ denotes extra states that are used to accommodate fractional occupations. \citep{kresse1994ab, Kresse_abinitio_MD, michaud2016rescu} It usually suffices to take $N_{x}$ to be about $5 - 10 \%$  of $N_e / 2$ while dealing with electronic temperatures up to a few thousand Kelvins. In this case, the occupation numbers associated with the states lying beyond the lowest $N_s$ can be conveniently set to $0$ without compromising the accuracy of the solution (ground-state energies and forces, for example) or aggravating SCF convergence.

The computation of the lowest $N_s$ eigenvalues $\{\epsilon_{i}\}_{i=1}^{N_s}$ and the corresponding eigenvectors $\{\psi_{i}\}_{i=1}^{N_s}$ of $H$ can be carried out through the use of direct or iterative eigensolvers. Subsequent to the computation of the eigenstates, the occupation fractions $\{f_i\}_{i=1}^{N_s}$ (with $0 \leq f_i \leq 1$) can be computed from the Fermi-Dirac function \citep{Mermin_Finite_Temp}
\begin{align}
f_{i} = f_F(\epsilon_i ),\;\text{with}\;f_F(\epsilon) = \frac{1}{1+\exp\big(\frac{\epsilon - \epsilon_F}{k_B\,\Theta_e}\big)}\,,
\label{eq:Fermi_function}
\end{align}
where $\Theta_e$ is the electronic temperature, $k_B$ denotes the Boltzmann constant, and the Fermi level $\epsilon_F$ can be determined by solving the constraint equation
\begin{align}
2\sum_{i=1}^{N_s} f_i = N_e\,.
\label{eq:fermi_constraint}
\end{align}
The use of fractional occupation (also known as smearing) \citep{kresse1994ab, Kresse_abinitio_MD, aarons2016perspective} allows us to overcome numerical difficulties associated with 
possible degeneracy of eigenstates near $\epsilon_F$.

Using the results from the above computations, the (discretized) density matrix (also referred to as the Fermi matrix at finite electronic temperature) of the system can be calculated. This $N_b\times N_b$ sized 
matrix is defined as 
\begin{align}
P = f_F(H)\,,
\label{eq:DM_Fermi}
\end{align}
and using the fact that $f(\epsilon_i) = 0$ for $i > N_s$, it can be rewritten using the eigenvectors of $H$ as 
\begin{align}
\label{eq:DM_eigenvectors}
P = \sum_{i=1}^{N_s} f_i\,\psi_{i}\,\psi_{i}^T\,.
\end{align}
Denoting the collection of the eigenvectors $\{\psi_{i}\}_{i=1}^{N_s}$ as the $N_b \times N_s$ matrix $X$, and the $N_s \times N_s$ diagonal matrix of occupation numbers as $\mathfrak{F}$ (i.e., $\mathfrak{F}_{i,i} = f_i$ for $i=1,\ldots,N_s$), a more compact matrix form of the above expression (i.e., Eq.~\ref{eq:DM_eigenvectors}) is 
\begin{align}
P = X\,\mathfrak{F}\,X^T\,.
\label{eq:DM_matrix_form}
\end{align}
The matrix $P$ contains all the information required for progressing with the SCF iterations --- in particular, if the basis functions used for the discretization are denoted as $\big\{e_j(x)\big\}_{j=1}^{N_b}$, then the real-space electron density can be expressed\footnote{Alternately, the eigenvectors can be  expressed in real space as $\psi_i(x) = \sum_{j=1}^{N_b} \psi_{i,j}\,e_j(x)$, whereupon the electron density can be obtained as $\rho(x) = 2 \sum_{i=1}^{N_s}\,f_i\,\big[\psi_i(x)\big]^2$. This approach is conventionally used, for example, in planewave and other such codes\citep{banerjee2015spectral} employing  global basis sets, but is not as suitable in the context of localized basis sets \citep{banerjee2016chebyshev}, as we employ here.} using the matrix entries of $P$ as 
\begin{align}
\rho(x) = 2 \sum_{j=1}^{N_b}\sum_{j'=1}^{N_b}P_{j,j'}\,e_j(x)\,e_{j'}(x)\,.
\label{eq:rho_from_DM}
\end{align}
In the process of computing the density matrix $P$, it is often simpler to compute an alternate set of orthonormal vectors $\{\phi_i\}_{i=1}^{N_s}$ that span the same subspace as the eigenvectors (i.e., the occupied subspace). If the collection of these alternate vectors is expressed as an $N_b \times N_s$ matrix $Y$, there must exist an orthogonal $N_s \times N_s$ matrix $Q$ such that $X = Y Q$, and Eq.~\eqref{eq:DM_matrix_form} then takes the form 
\begin{align}
\label{eq:proj_DM}
P = Y\,\big(Q\,\mathfrak{F}\,Q^T\big)\,Y^T\,.
\end{align}
The $N_s \times N_s$ matrix $\tilde{P} = Q\,\mathfrak{F}\,Q^T$ will be
referred to as the \emph{projected density matrix}.
Equation~\eqref{eq:proj_DM} indicates that the density matrix $P$ may be computed using alternative vectors $\{\phi_i\}_{i=1}^{N_s}$ if the projected density matrix is available along with the vectors $\{\phi_i\}_{i=1}^{N_s}$.

A straightforward way of computing such a set of alternate orthonormal vectors is through the use of Chebyshev polynomial filtering followed by explicit orthonormalization. \citep{Saad_large_eigenvalue_book} Specifically, we may start with a $N_b \times N_s$ block of linearly independent vectors $Y_0$, and apply a Chebyshev polynomial filter matrix $p_m(H)$ to $Y_0$. The filter polynomial $p_m(\cdot)$ can be specifically scaled and a sufficiently high filter order $m$ can be chosen so that the eigenvectors $\{\psi_{i}\}_{i=1}^{N_s}$ are amplified in the resulting filtered vectors $Y_1 = p_m(H)\,Y_0$. \citep{Serial_Chebyshev, Parallel_Chebyshev, zhou_2014_chebyshev} To avoid linear dependencies, we may then orthonormalize the vector block $Y_1$. The resulting set of orthonormal vectors will (approximately) span the occupied subspace. This  strategy has been combined with subspace iteration techniques for use in various electronic structure codes\citep{Serial_Chebyshev, Parallel_Chebyshev, banerjee2015spectral, banerjee2016chebyshev, ghosh2017sparc,ghosh2017sparc1,CyclicDFT_JMPS} and it can successfully deal with metallic as well as insulating systems.

We note that a special situation arises when the system in question is an insulator. In this case, the matrix of occupation numbers $\mathfrak{F}$ is the identity matrix, and so Eq.~\ref{eq:proj_DM} reduces to 
\begin{align}
\label{eq:proj_DM_insulator}
P = Y\,Y^T\,,
\end{align}
as $Q$ is an orthogonal matrix. Thus, for an insulating system, we only
need a way of computing a set of orthonormal vectors that span the
occupied subspace in order to compute the density matrix $P$. For a
metallic system, additional work is needed to compute the projected density matrix. 

\subsubsection{Direct computation of the projected density matrix}
\label{subsubsec:direct_proj_DM}
Considering the expression for the projected density matrix $\tilde{P} =
Q\,\mathfrak{F}\,Q^T$, we see that the evaluation of this expression
requires the computation of the occupation numbers $\displaystyle\{f_i = f_F(\epsilon_i)\}_{i=1}^{N_s}$ as well as the matrix $Q$. These quantities can be computed by carrying out an eigenvalue decomposition of the \emph{projected Hamiltonian matrix}, i.e., the $N_s \times N_s$ matrix $\tilde{H} = Y^T\,H\,Y$. The occupation numbers can be computed using the eigenvalues of $\tilde{H}$ since these are the same as the lowest $N_s$ eigenvalues of $H$ (i.e., $\displaystyle\{\epsilon_i\}_{i=1}^{N_s}$). Furthermore, the  eigenvectors of $\tilde{H}$ are the columns of the matrix $Q$. 

To verify this, we first write down the eigendecomposition of $H$ (for the lowest $N_s$ states) as $H\,X = X\,\Lambda$. Next, using $X = Y\,Q$, we get 
\begin{align}
H\,Y\,Q = Y\,Q\,\Lambda\,.
\label{eq:intermediate_calc_1}
\end{align}
Premultiplying with $Y^T$ and using the orthonormality of the column vectors in $Y$, we get 
\begin{align}
\big(Y^T\,H\,Y\big)\,Q = \tilde{H}\,Q = Q\,\Lambda.
\label{eq:intermediate_calc_2}
\end{align} 
Additionally, the expressions $\tilde{P} =  Q\,\mathfrak{F}\,Q^T$ and $\tilde{H} =  Q\,\Lambda\,Q^T$ allow us to interpret the projected density matrix in terms of the projected Hamiltonian matrix as $\tilde{P} = f_F(\tilde{H})$. 

In the context of Krylov subspace projection methods,
\citep{Saad_large_eigenvalue_book, parlett1998symmetric} the steps involving
the construction of the projected Hamiltonian matrix, 
computation of its eigendecomposition, and computation of (approximate) eigenvectors of $H$ using the expression $X = Y\,Q$ constitute the Rayleigh-Ritz process, 
wherein the last step corresponds to subspace rotation. 
As the system size grows, the dimension of the Hamiltonian $N_b$ and number of states $N_s$ grow as well. Correspondingly, the eigendecomposition and subspace rotation steps begin to consume more and more computation time, thus making it infeasible to directly compute the projected density matrix as described above. Instead, a complementary subspace strategy may be formulated to mitigate these issues, as we now describe.

\subsubsection{Complementary subspace computation of the projected density matrix}
In light of the above discussion, it appears that for a generic system
(i.e., one with some degree of fractional occupation), we require all the $N_s$ eigenstates of
$\tilde{H}$ to be computed on every SCF iteration in order 
to progress with the SCF iterations. A crucial observation, however, is
that for electronic temperatures typically encountered in practice 
(e.g., $\Theta_e \lesssim 3000$ Kelvin), a large majority
of the occupation numbers $\{f_i\}_{i=1}^{N_s}$ are equal to $1$. We
denote states $1$ through $N_1$ as those with occupation numbers equal
to $1$. The remaining states, from $N_{1}+1$ through $N_s$,
have occupations numbers less than $1$. Let $N_t$ be the
number of these fractionally occupied states, i.e., $N_t = N_s - N_1$. Denoting the eigenvectors of the projected density matrix as $\{\tilde{\psi}_i\}_{i=1}^{N_s}$ (the columns of the matrix $Q$), we may rewrite the expression for the projected density matrix as 
\begin{align}
\nonumber
\tilde{P} &= \sum_{i=1}^{N_s} f_i\,\tilde{\psi}_{i}\,\tilde{\psi}_{i}^T\,\\\nonumber
%&= \sum_{i=1}^{N_1} f_i\,\tilde{\psi}_{i}\,\tilde{\psi}_{i}^T + \sum_{i=N_1 + 1}^{N_s} f_i\,\tilde{\psi}_{i}\,\tilde{\psi}_{i}^T\,\\\nonumber
&=\sum_{i=1}^{N_1} \tilde{\psi}_{i}\,\tilde{\psi}_{i}^T + \sum_{i=N_1 + 1}^{N_s} f_i\,\tilde{\psi}_{i}\,\tilde{\psi}_{i}^T\,\\\nonumber
&=\sum_{i=1}^{N_1}\!\tilde{\psi}_{i}\,\tilde{\psi}_{i}^T + \sum_{i=N_1 + 1}^{N_s}\!\tilde{\psi}_{i}\,\tilde{\psi}_{i}^T- \sum_{i=N_1 + 1}^{N_s}\!\tilde{\psi}_{i}\,\tilde{\psi}_{i}^T + \sum_{i=N_1 + 1}^{N_s} \!f_i\,\tilde{\psi}_{i}\,\tilde{\psi}_{i}^T\,\\\label{eq:CS_DM_1}
&=\sum_{i=1}^{N_s} \tilde{\psi}_{i}\,\tilde{\psi}_{i}^T - \sum_{i=N_1 + 1}^{N_s} \!\tilde{\psi}_{i}\,\tilde{\psi}_{i}^T+ \sum_{i=N_1 + 1}^{N_s} f_i\,\tilde{\psi}_{i}\,\tilde{\psi}_{i}^T\\ 
&=\tilde{\calI} - \sum_{i=N_1 + 1}^{N_s} (1 - f_i)\,\tilde{\psi}_{i}\,\tilde{\psi}_{i}^T\,.
\label{eq:CS_DM_2}
\end{align}
In Eq.~\ref{eq:CS_DM_2} above, $\tilde{\calI}$ denotes the identity matrix of dimension $N_s \times N_s$. That the first term of Eq.~\ref{eq:CS_DM_1} is the identity matrix follows from the fact that the vectors $\{\tilde{\psi_i}\}_{i=1}^{N_s}$ are the eigenvectors of $\tilde{H}$, a symmetric matrix, and so form a resolution of the identity.

The above expression suggests that if the $N_t$ top eigenvectors $\tilde{\psi_i}$ and corresponding occupation numbers $f_i$ are known, the projected density matrix $\tilde{P}$ may be computed. Thus, instead of determining the full $N_s \times N_s$ set of vectors, we need to determine only an extremal block of vectors (of dimension $N_s \times N_t$), corresponding to the states $i = N_1+1$ to $N_s$. 

To compute the corresponding occupation numbers, we rewrite the equation $\displaystyle 2 \sum_{i=1}^{N_s}f_i = N_e$ as 
\begin{align}
2 \big(\sum_{i=1}^{N_1}f_i +  \sum_{i=N_1 + 1}^{N_s} f_i \big)= N_e \implies \sum_{i=N_1 + 1}^{N_s} f_i = N_e / 2 - N_1.
\label{eq:Fermi_Constraint_in_CS}
\end{align}
The above algebraic equation may be solved for the Fermi level $\epsilon_F$ and occupation numbers $\{f_i\}_{i=N_1+1}^{N_s}$. 

Once the projected density matrix $\tilde{P}$ has been obtained, the full (i.e., $N_b \times N_b$) density matrix can be obtained as $P = Y \tilde{P} Y^{T}$. To simplify this further, let us denote by $\tilde{C}_{\mathfrak{F}}$ the $N_s \times N_t$ matrix consisting of the vectors $\{\sqrt{1-f_i}\,\tilde{\psi}_i\}_{i=1}^{N_t}$ (i.e., each of the top eigenvectors of $\tilde{H}$ scaled by the quantity $\sqrt{1-f_i}$\,). Then Eq.\ref{eq:CS_DM_2} can be written in terms of $\tilde{C}_{\mathfrak{F}}$ as 
\begin{align}
\tilde{P} = \tilde{I} - \tilde{C}_{\mathfrak{F}}\tilde{C}^T_{\mathfrak{F}}\, 
\label{eq:Ptilde_CS_expression}
\end{align}
whereupon we obtain 
\begin{align}
\nonumber
P &= Y \tilde{P}Y^{T}= Y\,Y^{T} - Y\tilde{C}_{\mathfrak{F}}\,\tilde{C}_{\mathfrak{F}}^T\,Y^T\\
&=  Y Y^{T} - (Y\tilde{C}_{\mathfrak{F}})(Y\tilde{C}_{\mathfrak{F}})^T\,.
\label{eq:P_CS_expression}
\end{align}
As we explain later, this expression is particularly easy to evaluate in a localized basis set. In particular, within the DGDFT code, evaluation of the diagonal portions of the above expression can be carried out in a manner that avoids interprocess communication. This is sufficient for evaluating the real-space electron density and proceeding with SCF iterations.

Ground-state KS-DFT calculations also require computation of the band
energy $\displaystyle E_{b} = \sum_{i=1}^{N_s} f_i\,\epsilon_i$. Since
only the fractionally occupied states are 
in the complementary subspace scheme, we rewrite this as 
\begin{align}
\nonumber
E_{b} &= 2\,\bigg(\sum_{i=1}^{N_s} f_i \,\epsilon_i \bigg) = 2\,\bigg(\sum_{i=1}^{N_1} f_i \,\epsilon_i + \sum_{i=N_1+1}^{N_s} f_i \,\epsilon_i  \bigg)\\\nonumber
&= 2\,\bigg(\sum_{i=1}^{N_1} \epsilon_i + \sum_{i=N_1+1}^{N_s} f_i \,\epsilon_i  \bigg) = 2\bigg(\sum_{i=1}^{N_s} \epsilon_i + \sum_{i=N_1+1}^{N_s} f_i \,\epsilon_i  - \sum_{i=N_1+1}^{N_s} \epsilon_i \bigg)\\
&=  2\bigg(\text{Tr}(\tilde{H}) - \sum_{i=N_1+1}^{N_s} (1 - f_i)\,\epsilon_i \bigg)\,.
\end{align}
Thus, the trace of the projected Hamiltonian matrix, the top eigenvalues, and their corresponding occupation numbers are sufficient to compute the band energy. 

The electronic entropy can be obtained from just the fractionally occupied states as well. The electronic entropy is given by  
\begin{align}
\label{eq:e_entropy}
S = - 2 k_B \displaystyle \sum_{i=1}^{N_{s}}f_i\log{f_i} + (1-f_i)\,\log{(1-f_i)}\,.
\end{align}
By inspection, we see that the contribution of a state $i$ to the electronic entropy goes to zero as the occupation number for that state $f_i$ goes to $0$ or $1$. Hence, within the complementary subspace scheme, only the contribution of the fractionally occupied states is considered. This allows the simplification of Eq.~\ref{eq:e_entropy} to 
\begin{align}
\label{eq:e_entropy_CS}
S = -2 k_B \displaystyle \sum_{i=N_1+1}^{N_{s}}f_i\log{f_i} + (1-f_i)\,\log{(1-f_i)}\,,
\end{align}
which can be readily computed.

\subsection{Computation of top states}
We now discuss strategies for computing the $N_t$ topmost occupied states of the projected Hamiltonian matrix $\tilde{H}$. This is the key step in the CS2CF methodology. To the extent that the top states of the projected Hamiltonian can be obtained more quickly than all states, the methodology will outperform standard dense and sparse-direct solvers which obtain all states.

To obtain the top states as efficiently as possible, we exploit two key properties of the projected Hamiltonian. First, by construction (i.e., projection onto the occupied subspace) the projected Hamiltonian has quite limited spectral width, with maximum eigenvalue limited to that of the highest occupied state. Second, since we seek only the top few (typically $\leq 10 \%$) states, we have an extremal eigenvalue problem for a relatively small fraction of the spectrum.

The above properties suggest that iterative solution methods stand to be efficient to obtain the desired states. We explored two such approaches: (1) the LOBPCG method \citep{LOBPCG_1, LOBPCG_3, deursch2017robust}, as employed in Ref.~\citep{michaud2016rescu}, and (2) CheFSI.

\subsubsection{Use of LOBPCG}
\label{subsubsec:lobpcg_top_states}
We first implemented an unpreconditioned version of the LOBPCG method \citep{LOBPCG_1, LOBPCG_3, deursch2017robust}. 
The topmost $N_t$ states of $\tilde{H}$ were obtained by computing the bottommost $N_t$ states of $-\tilde{H}$. The initial vectors for the LOBPCG iterations were obtained from the results of a direct  diagonalization of $\tilde{H}$ (using LAPACK / ScaLAPACK) from a previous SCF step. Computation of matrix-vector products was carried out directly by the use of dense linear algebra (BLAS) routines. 

Overall, this strategy works reasonably well in practice. When compared against results from the direct  diagonalization of $\tilde{H}$, 
a few iterations of LOBPCG are typically enough to obtain the top eigenstates to desired accuracy at a fraction of the cost. However, as the system size increases, so does the total number of states $N_s$ and number of top states $N_t$. Under these circumstances, the well known computational bottlenecks of the LOBPCG algorithm associated with dense linear algebra operations begin to become apparent. Replacing serial dense linear algebra operations in LOBPCG with the corresponding parallel versions (i.e., PBLAS and ScaLAPACK routines \citep{choi1995proposal, choi1995scalapack, ScaLAPACK_1}) did not 
significantly improve performance 
since the computational bottlenecks of LOBPCG 
also suffer from scalability issues. We therefore turned to a different strategy, as we describe below.
\subsubsection{Use of CheFSI --- two-level polynomial filtering strategy}
\label{subsubsec:chefsi_top_states}
To mitigate the aforementioned issues, we replaced the LOBPCG algorithm with Chebyshev polynomial filtered subspace iteration (CheFSI). This turns the overall iterative strategy into one that employs two levels of Chebyshev polynomial filtering on every SCF step. The first (or outer) level allows the computation of a set of orthonormal vectors that (approximately) span the occupied subspace of $H$. The second (or inner) level uses CheFSI 
to compute the $N_t$ topmost states of $\tilde{H}$, or equivalently, the $N_t$ lowest states of $-\tilde{H}$. 

This turns out to be a much more effective strategy for a number of reasons.
First, by virtue of the limited spectral width of the projected Hamiltonian, 
a low-order polynomial filter suffices for the inner CheFSI iterations. In fact, for all calculations reported here, we found a filter order of $4$ or lower to be sufficient. 
Secondly, as explained in previous work \citep{zhou_2014_chebyshev}, depending on the initial guess provided, as well as the spectral width of the matrix, the use of CheFSI to determine eigenstates of sufficient accuracy often requires the application of multiple CheFSI cycles. These factors appear to work in our favor and we found that $5$ or fewer CheFSI cycles were sufficient in all cases considered, provided that the starting vectors for the inner CheFSI iteration were obtained using results from the previous SCF step. Finally, a significant fraction of the time involved in the inner CheFSI iteration is spent on evaluation of the Chebyshev polynomial filter $p_{\widetilde{m}}(-\tilde{H})$ as applied to an $N_s \times N_t$ block of vectors. This operation is based on matrix-matrix multiplications (i.e., GEMM operations in BLAS) and it parallelizes quite efficiently when PBLAS routines \citep{choi1995proposal} are used. Hence, the scalability of the inner CheFSI operation turns out to be more favorable as compared to LOBPCG. 

It is worthwhile to discuss the computational complexity of the
above procedure for determining the top states of $\tilde{H}$. If an
inner Chebyshev filter of order $\widetilde{m}$ is employed, the
computational cost of applying the Chebyshev filter is
$O(\widetilde{m}\,N_s^2\,N_t)$. 
The subsequent orthonormalization and projection, diagonalization, and rotation steps
associated with the inner problem incur costs of $O(N_s\,N_t^2 +
N_t^3)$,  $O(N_t^3)$, and $O(\,N_s\,N_t^2)$, respectively, leading to an overall cost of
$O(\widetilde{m}\,N_s^2\,N_t + N_s\,N_t^2 + N_t^3)$
for each inner CheFSI cycle. This estimate makes clear that
it is advantageous to reduce $N_t$ as much as possible in practical calculations using the CS2CF strategy, as long as this reduction does not adversely affect the accuracy or convergence of the calculation. As described later, this can be done in some cases based on numerical arguments (e.g., there is no need to account for fractionally occupied states which do not affect energies and forces appreciably) or physical ones (certain systems might be expected to have only a few fractionally occupied states, based on symmetry arguments, for example).

If $\widetilde{m}$ is small (as in practical calculations),
the above estimate appears to be of lower complexity than 
the $O(N_s^3)$ cost associated with the direct diagonalization
of the projected Hamiltonian matrix.  However, noting that $N_t$ is a small
fraction of $N_s$ (typically less than 10\%), we see that
the asymptotic complexity associated with the inner CheFSI procedure is
the same as that of direct diagonalization of $\tilde{H}$ (i.e.,
$O(N_s^3)$) but with a lower prefactor. As we show in Section \ref{sec:Results}, this lower prefactor does indeed result in significantly lower computational wall times when the inner CheFSI technique is used in lieu of explicit diagonalization of $\tilde{H}$. 

\subsection{Implementation}
\label{subsec:DGDFT_Implementation}
Eq.~\ref{eq:P_CS_expression} suggests that for the success of the
complementary subspace strategy, it is essential to be able to compute
the full density matrix $P$ in an efficient manner once the vector
blocks $Y$ and $C_{\mathfrak{F}}$ are available. As described above, in
the two-level CheFSI scheme, the outer Chebyshev polynomial filtering
iterations allow us to compute the vector block $Y$ (using $H$) while
the inner CheFSI iterations allow us to compute the vector block
$C_{\mathfrak{F}}$ (using $\tilde{H}$). Using these computed quantities,
evaluating Eq.~\ref{eq:P_CS_expression} naively would incur a
computational cost of $O(N_{b}^2N_s + N_bN_sN_t + N_{b}^2N_t)$. However,
if the basis set used for the discretization is strictly localized, the
computation of certain entries of the density matrix $P$ can be avoided
during the SCF iterations, thus resulting in significant reductions in
computational cost. Specifically, according to Eq.~\ref{eq:rho_from_DM}, if the basis functions $e_{j}(x)$ and $e_{j'}(x)$ have nonoverlapping support, then it is redundant to compute the density matrix entry $P_{j,j'}$ since this term does not contribute to the real-space electron density. With this observation and a few additional factors (as detailed  below)  in mind, we have implemented the two-level CheFSI based complementary subspace strategy within the framework of the discontinuuos Galerkin electronic structure method (specifically, the DGDFT code), as we now describe.

\subsubsection{Background on discontinuous Galerkin electronic structure method and DGDFT code}
The discontinuous Galerkin (DG) electronic structure method employs an adaptive local basis (ALB) set to solve the equations of KS-DFT in a discontinuous Galerkin framework. \citep{lin2012adaptive, zhang2017adaptive} The methodology has been implemented in the Discontinuous Galerkin Density Functional Theory (DGDFT) code for large-scale parallel electronic structure calculations. \citep{hu2015dgdft, banerjee2016chebyshev}  The DGDFT approach to solving the KS-DFT equations involves partitioning the global simulation domain into a set of subdomains (or elements). The Kohn-Sham equations are then solved locally in and around each element. These local calculations are used to generate the ALBs (in each element) and the Kohn-Sham equations in the global simulation domain are then discretized using them. The ALBs form a discontinuous basis set globally with discontinuities occurring at the element boundaries. Subsequent to the generation of the ALBs, the interior penalty discontinuous Galerkin approach~\cite{Arnold1982} is used for constructing the Hamiltonian matrix. This formulation ensures that the global continuity of the relevant Kohn-Sham eigenstates and related quantities such as the electron density is sufficiently maintained. 

As the number of ALBs is increased, the solution obtained by the above procedure converges systematically to the infinite basis set limit. Since the ALBs incorporate local materials physics into the basis, an efficient discretization of the Kohn-Sham equations can be obtained in which chemical accuracy in total energies and forces can be attained with a few tens of basis functions per atom. \citep{lin2012adaptive, zhang2017adaptive} Additionally, the rigorous mathematical foundations of the discontinuous Galerkin method allow the errors in the above approach to be systematically gauged by means of \textit{a posteriori} error estimators. \citep{dgdft_posteriori,lin2015posteriori,lin2016posteriori} Thus, DGDFT combines the key advantage of planewave basis sets in terms of systematic improvability with that of localized basis sets in reducing basis set size. The DG framework for solution of the Kohn-Sham equations (as implemented in the DGDFT code) has been successfully used to study complex materials problems involving many thousands of atoms. \citep{banerjee2016chebyshev, hu2015edge, hu2015dgdft}

Despite the many successes of DGDFT in studying a wide variety of large-scale materials problems, a persistent issue has been to obtain the electron density from the discretized Kohn-Sham Hamiltonian in an efficient and scalable manner for large systems (i.e., systems containing a thousand or more atoms). To address this issue, we have recently investigated the use of Chebyshev polynomial filtered subspace iteration (CheFSI) within DGDFT. \citep{banerjee2016chebyshev} While this technique has the same asymptotic computational complexity as traditional diagonalization based methods (i.e., $O(N_s^3)$, with $N_s$ denoting the number of Kohn-Sham states), it has a substantially lower prefactor compared to the existing alternatives (based on direct diagonalization using ScaLAPACK, for instance) within DGDFT. This stems from several favorable properties of the discretized Hamiltonian matrix in DGDFT. These include: a small dimension (e.g., a few tens times the number of atoms) which leads to lower linear algebra operation costs, a relatively low spectral width which ensures Chebyshev polynomials of relatively low order can be employed, and finally, an underlying block sparse structure which ensures that matrix vector products can be carried out with high computational efficiency. These features, along with the favorable parallel scalability of the DG-CheFSI approach have allowed us to tackle systems containing several thousands of atoms in minutes of wall time per SCF step on large-scale computational platforms. \citep{banerjee2016chebyshev} 

Our experience has shown that the limiting computational bottleneck in such large-scale calculations using the DG-CheFSI approach turns out to be associated with the subspace diagonalization and subspace rotation steps of the Rayleigh-Ritz process. \citep{banerjee2016chebyshev} In light of this observation, we view the current contribution as one which directly confronts the computational bottlenecks associated with the above steps and replaces them with the complementary subspace strategy based on an inner level of Chebyshev filtering. In particular, the factors which lead to the success of the DG-CheFSI approach (i.e., favorable properties of the discretized Hamiltonian matrix, good parallel scalability of various operations, etc.) also ensure that the two-level CheFSI based complementary subspace strategy performs with great efficiency when implemented within the DGDFT framework. We now outline some specific implementation details of the above strategy within the DGDFT code.

\subsubsection{Implementation details}
\label{subsubsec:implementation_details}
We highlight a few important details of the implementation of the two-level CheFSI based complementary subspace strategy within DGDFT. The first concerns the manner in which the code transitions from regular CheFSI based SCF iterations to the use of the complementary subspace strategy. As discussed in a previous section, the use of iterative solvers to evaluate the topmost states of the projected Hamiltonian $\tilde{H}$ requires good initial approximations in order to avoid excessive iterations. Consequently, in static (i.e., fixed atomic positions) Kohn-Sham calculations, we first carry out about $4$--$5$ SCF iterations using the conventional CheFSI technique. The eigenstates of $\tilde{H}$ are computed directly by use of LAPACK or ScaLAPACK routines during these iterations. The $N_t$ top eigenvectors from the last conventional CheFSI iteration are subsequently used as the initial guess for the iterative solvers in the first complementary subspace based SCF iteration. Furthermore, in case of the two-level CheFSI strategy, the bounds for the inner-level Chebyshev polynomial filter (i.e., the one used to compute the topmost states of $\tilde{H}$) are computed using the eigenvalues evaluated in the previous (conventional CheFSI based) SCF step. Following this transition, the topmost eigenvalues and eigenvectors of $\tilde{H}$ are always stored between SCF iterations for use by the iterative solvers in subsequent SCF steps. During molecular dynamics or geometry optimization runs, we have used the above methodology only during the first ionic step. For subsequent ionic steps, the complementary subspace strategy is used exclusively in every SCF iteration.

The second detail pertains to the parallelization aspects of the two-level 
CheFSI based complementary subspace strategy. The parallelization
strategies involved in the first-level Chebyshev polynomial filter
computation (i.e., the one associated with $H$) are described in earlier
work\citep{banerjee2016chebyshev}.  Parallelization of the various
linear algebra operations associated with the second level of CheFSI
iterations are carried out with the use of PBLAS and ScaLAPACK routines
\citep{choi1995proposal, choi1995scalapack, ScaLAPACK_1}. Accordingly,
the various matrices involved in these computations are
redistributed over two dimensional block cyclic process grids. The
various redistribution and parallel storage format interconversion
routines (employing ScaLAPACK's \textsf{pdgemr2d} routine or otherwise)
did not consume more than one percent of the total time spent in the complementary subspace strategy, even for the largest systems considered here.

Finally, the third detail pertains to the computation of the density matrix in DGDFT by using Eq.~\ref{eq:P_CS_expression}. Since the supports of the ALBs are confined to individual DG elements, the Hamiltonian matrix in DGDFT enjoys a block-sparse structure in which nonzero contributions arise due to an element and only its nearest neighbors. \citep{banerjee2016chebyshev, hu2015dgdft} The (full) density matrix enjoys this structure as well. \citep{hu2015dgdft} As noted above, the real-space electron density that must be updated in each SCF iteration only accumulates contributions from density matrix entries that are associated with basis functions with overlapping support. These factors combined imply that \emph{only the diagonal blocks} of the density matrix are required in DGDFT when the complementary subspace strategy is 
used to update the density in each SCF iteration. 
This is a significant reduction in the number of operations relative to what a naive inspection of Eq.~\ref{eq:P_CS_expression} would suggest and it is one of the primary reasons for the success of the  present strategy within DGDFT.

The DGDFT code uses a two-level parallelization strategy implemented via
Message Passing Interface (MPI) to handle inter-process communication.
\citep{hu2015dgdft} At the coarse grained level, the parallelism is
based on domain decomposition and work is distributed among processors
by DG elements. Further, multiple processors are assigned to each
element to achieve the second, finer level of parallelism. The
observations made above imply that computation of the diagonal blocks of
the density matrix incurs no communication between processors associated with different elements as long as the matrix $\tilde{C}_{\mathfrak{F}}$ is available locally on the processors working on a given element. Our implementation makes use of this observation to achieve a good balance of memory storage requirements and parallel scalability of linear algebra operations while working with the matrix $C_{\mathfrak{F}}$. Once SCF convergence has been achieved, the full density matrix (or, more precisely, all the nonzero blocks) needs to be computed. The two-level parallelization strategy implemented in DGDFT ensures that this computation can be done efficiently in parallel with a small contribution to the overall wall time.

\begin{figure}
\centering
\begin{tikzpicture}[node distance=2cm, scale=1, every node/.style={scale=0.55}]
\node (start) [long_startstop] {Initialize $\rho, V_{\text{eff}}$, etc.\\ Set random initial guess for $Y$.};
\node (proc1) [long_process, below of=start, yshift=-0.2cm] {Compute ALBs locally on each DG element.};
\node (proc2) [long_process, below of=proc1, yshift=0.1cm] {Construct Hamiltonian matrix $H^{\text{DG}}$.};
\node (dec1) [decision, below of=proc2, yshift=-0.1cm] {First SCF iteration ?};
\node (proc3) [long_process, right of=dec1, xshift=4.5cm] {Perform 3 -- 4 standard CheFSI cycles on $Y$ using $H^{\text{DG}}$.};
\node (proc4) [long_process, below of=dec1, yshift=-0.3cm] {Align $Y$ with current basis (see ref \cite{banerjee2016chebyshev}).};
\node (proc5) [long_process, below of=proc4, yshift=-0.5cm] {Perform Chebyshev polynomial filtering on $Y$ using $H^{\text{DG}}$,\\ Orthonormalize columns of $Y$.};
\node (proc6) [long_process, below of=proc5, yshift=-0.7cm] {Compute projected Hamiltonian $\tilde{H} = Y^T\,H^{\text{DG}}\,Y$.};
\node (dec2) [decision, below of=proc6, yshift=-0.4cm] {Is SCF iteration < 5 ?};
\node (proc7) [long_process, right of=dec2, xshift=4.5 cm] {Perform subspace diagonalization and rotation steps. Save topmost $N_t$ eigenvectors of $\tilde{H}$.};
\node (proc8) [long_process, below of=proc7, yshift=-2.0 cm] {Compute Fermi level and occupation numbers for all eigenstates. Form diagonal blocks of density matrix $P$.};
\node (proc9) [very_long_process, below of=dec2, yshift=-1.3 cm] {Compute topmost $N_t$ eigenstates of $\tilde{H}$ using second-level CheFSI scheme.\\ Use saved top states as initial guess for this.};
\node (proc10) [very_long_process, below of=proc9, yshift=-2.3 cm] {Compute Fermi level using Eq.~\ref{eq:Fermi_Constraint_in_CS}. Compute top occupation numbers. Compute scaled top eigenvectors $\tilde{C}_{\mathfrak{F}}$ of $\tilde{H}$.\\ Form diagonal blocks of density matrix $P$ using Eq.~\ref{eq:P_CS_expression}.};
\node (proc11) [long_process, below of=proc10, yshift=-1.6 cm] {Compute real-space electron density $\rho(x)$. Update potentials.};
\node (dec3) [decision, below of=proc11, yshift=-0.3cm] {Is SCF converged ?};
\node (stop) [long_startstop, below of=dec3, yshift=-0.8 cm] {Compute full density matrix. Evaluate energies, forces, etc. Output results.};
\draw [arrow] (start) -- (proc1);
\draw [arrow] (proc1) -- (proc2);
\draw [arrow] (proc2) -- (dec1);
\draw [arrow] (dec1) -- node[anchor=south] {yes} (proc3);
\draw [arrow] (dec1) -- node[anchor=west] {no} (proc4);
\draw [arrow] (proc3) |- (proc5);
\draw [arrow] (proc4) -- (proc5);
\draw [arrow] (proc5) -- (proc6);
\draw [arrow] (proc6) -- (dec2);
\draw [arrow] (dec2) -- node[anchor=south] {yes} (proc7);
\draw [arrow] (dec2) -- node[anchor=west] {no} (proc9);
\draw [arrow] (proc7) -- (proc8);
\draw [arrow] (proc9) -- (proc10);
\draw [arrow] (proc8) |- (proc11);
\draw [arrow] (proc10) -- (proc11);
\draw [arrow] (proc11) -- (dec3);
%\draw [arrow] (proc5) -- (proc6);
%\draw [arrow] (proc6) -- (dec2);
\draw [arrow] (dec3) -- node[anchor=west] {yes} (stop);
\draw [arrow] (dec3) -| node[anchor=north] {no} ([xshift=-0.75cm]proc1.south west) |- (proc1);
\end{tikzpicture}
\caption{\footnotesize{Flowchart depicting the various steps of the CS2CF strategy within DGDFT for a  ground-state Kohn-Sham calculation. Note that $Y$ is an $N_b \times N_s$ block of orthonormal vectors that spans the occupied subspace of the Kohn-Sham Hamiltonian $H$ as shown in Eq.~\ref{eq:proj_DM}. $H^{\text{DG}}$ denotes the Hamiltonian matrix in the adpative local basis set used in DGDFT. 
}} \label{fig:flowchart}
\end{figure}
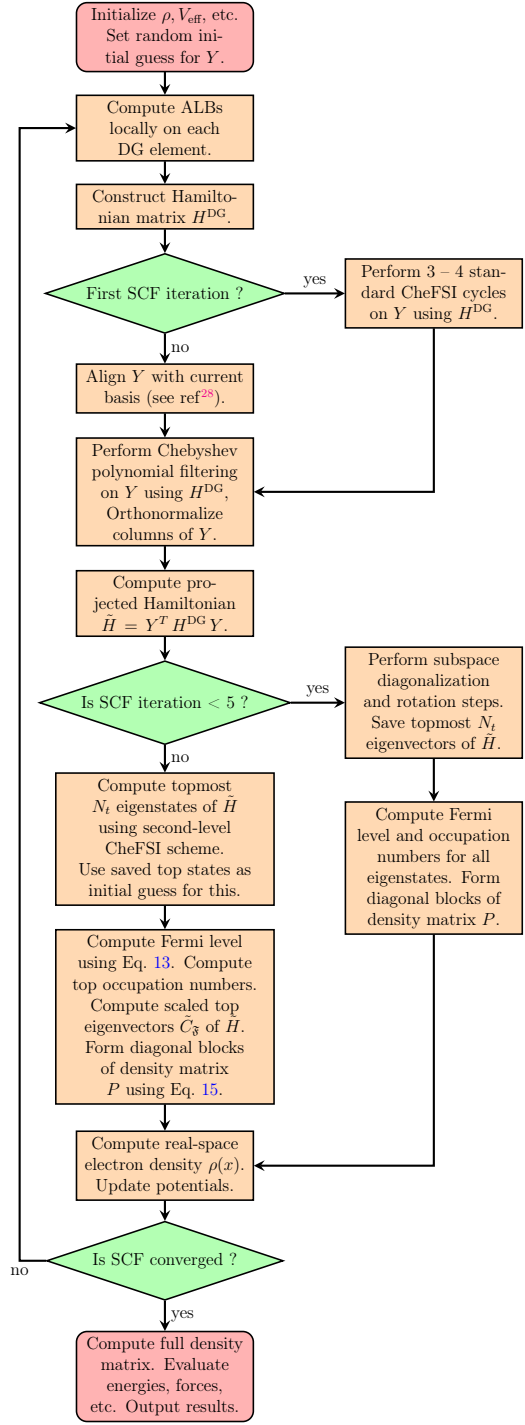

Figure \ref{fig:flowchart} depicts the various steps of the CS2CF strategy within DGDFT for a static ground-state calculation. {Table \ref{tab:parameters} summarizes the values of the various parameters used within the strategy for such calculations.}
\begin{table}
\begin{center}
\scriptsize{
{
\begin{tabular}{| c | c | c |}\hline
& & \\
Parameter & Criteria used for selecting & Value commonly used \\
& parameter value & in this work \\ & &\\\hline
& &\\
Total no.~of electronic  & Number of electrons $N_e$ in system, & $1.05 \times N_e / 2$\\
states ($N_s$) & type of system (metal/insulator, etc.) &   \\ & & \\\hline
& &\\
No.~of top states ($N_t$) & Type of system (metal/insulator, etc.) & $0.1 \times N_s$\\
& & \\\hline
& &\\
Outer Chebyshev polynomial & Spectral width of Hamiltonian $H$ & $30 - 50$\\
filter order & (depends on type of atoms in system) & \\ & & \\\hline
& &\\
Inner Chebyshev polynomial & Spectral width of  projected Hamiltonian $\tilde{H}$, & $4$\\
filter order & density of fractionally occupied states  &  \\ & & \\\hline
& &\\
No.~of inner CheFSI & Spectral width of projected Hamiltonian $\tilde{H}$, & $4$\\
cycles (using $\tilde{H}$) & density of fractionally occupied states &  \\ & & \\\hline
& &\\
No.~of initial SCF steps & Type of system (metal/insulator, etc.), & $4 - 5$\\
using regular CheFSI & nature of SCF convergence &  \\ & & \\\hline
\end{tabular}
}
}
\end{center}
\caption{\footnotesize{{Values of various parameters used for the CS2CF strategy in ground-state calculations at typical electronic temperatures. Note that, as suggested in previous work\citep{banerjee2016chebyshev, zhou_2014_chebyshev}, the first SCF step of a ground-state calculation employs multliple (regular) CheFSI cycles (typically, $3-4$) while starting from randomly initialized wavefunctions.}}}
\label{tab:parameters}
\end{table}

\section{Results and discussion}
\label{sec:Results}
To demonstrate the accuracy, efficiency, and parallel scaling of the CS2CF methodology, 
we apply it to five prototypical systems 
encompassing metals, semimetals, semiconductors, and insulators,
ranging is size from a few hundred atoms to over 27,000.
The fundamental unit cells (i.e., atomic configurations replicated to generate the various large systems examined subsequently) of these systems are summarized in Table~\ref{table:systems}. 
The first system, referred to as \emph{Electrolyte3D}, consists of a three-dimensional bulk lithium-ion electrolyte system originating from the design of energy storage devices.  Atoms of hydrogen, lithium, carbon, phosphorus, oxygen, and fluorine, numbering $318$ in total, are present in a single unit cell of this system. It serves as a protoypical bulk disordered insulating system. 
The second, referred to as \emph{SiDiamond3D}, consists of atoms of
crystalline silicon in the diamond structure, with $8$ atoms in
the unit cell. Silicon is a well known semiconductor and, in its
crystalline form, has an LDA band gap of $\sim 0.6$ eV. Thus, it
tends to have a small number of fractionally occupied states in 
Kohn-Sham calculations at room temperature. 
The third system, referred to as \emph{Graphene2D}, consists of a sheet of graphene
for which the unit cell contains 180 carbon atoms. This serves as a 
prototype for a two-dimensional semimetallic system. 
The fourth
system consists of atoms of lithium in a body centered cubic
configuration with $16$ atoms in the unit cell. We will refer to this
system as \emph{LiBCC3D}. 
Finally, the fifth system consists of atoms of
copper in a face centered cubic configuration with 4 atoms in the unit
cell. We will refer to this system as \emph{CuFCC3D}. The LiBCC3D and
CuFCC3D systems serve as prototypical examples of simple and more
complex bulk metallic systems, respectively. 
To remove periodicities in larger cells produced by replication,
we added mild random
perturbations to the atomic positions for all the crystalline/periodic 
systems mentioned above, before using them in calculations.

Together, these five systems were chosen for their technological relevance as well as the fact that KS-DFT calculations on large supercells based on these can be challenging. Additionally, the electronic properties of these systems cover a broad spectrum -- this has helped us ensure that the computational strategy presented in this work is able to deal successfully with different kinds of materials systems, without any computational difficulties arising from the physical nature of the system. 
\begin{table}
\begin{center}
{\tiny
\begin{tabularx}{\textwidth}{|c|c|c|c|c|c|C|}
 \hline
 & & & & & &\\
 {{System}} & {{Type}} & {No. of atoms} & {Types of atoms} & {No. of electrons} & {ALBs per atom} & {Representative image}\\
 {{}} & {{}} & {in unit cell} & {(elements)} & {in unit cell} & {in DGDFT} & {of unit cell}\\& & & & & &\\\hline
   & & & & & &\\ 
   Electrolyte3D & Bulk, Insulating & $318$ & C,H,F,Li,O & $1104$ & $40$ & \includegraphics[scale=0.12]{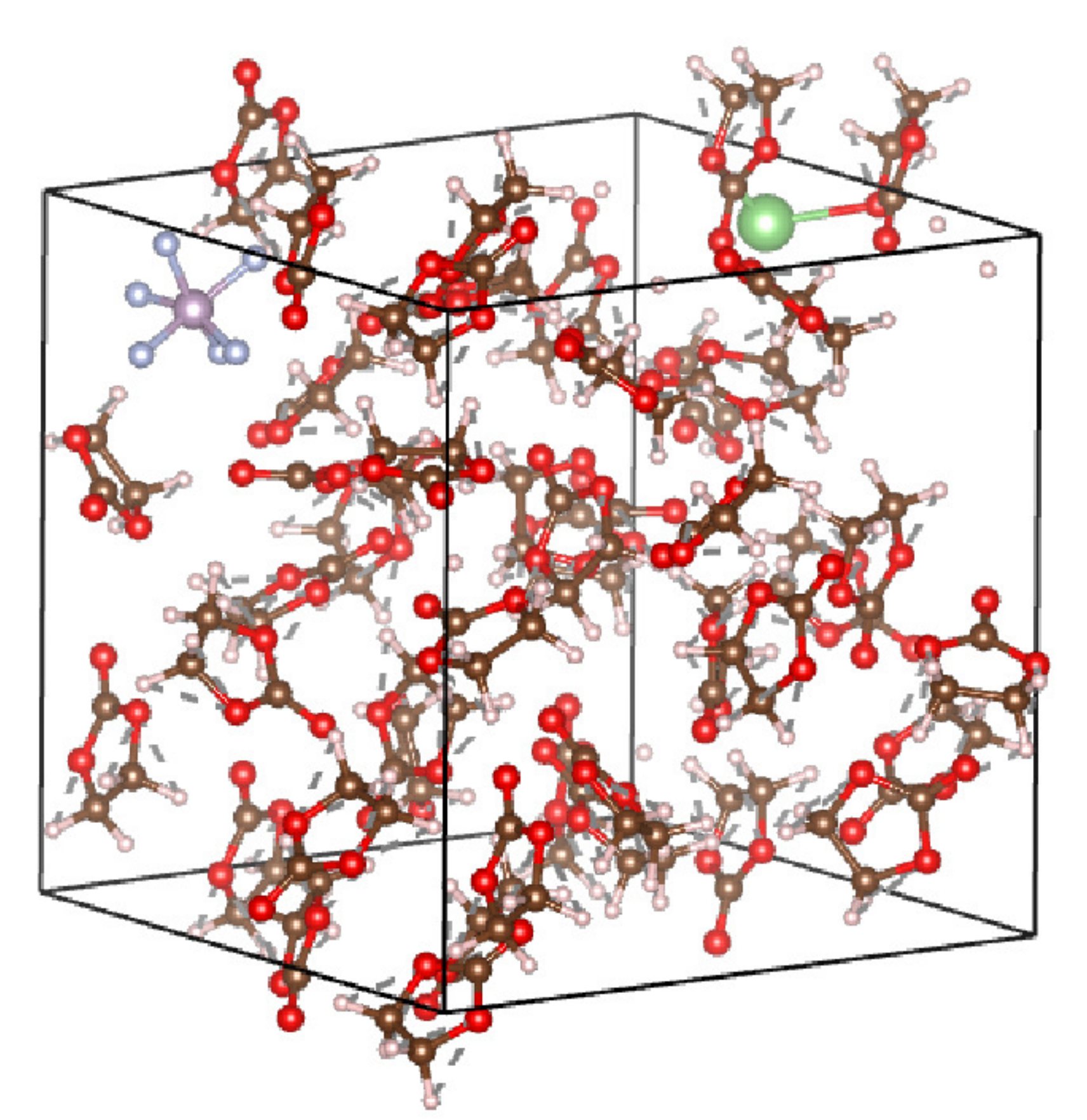} \\\hline
      & & & & & &\\
  SiDiamond3D  & Bulk, Semiconducting & $8$ & Si & $16$ & $40$ & \includegraphics[scale=0.1]{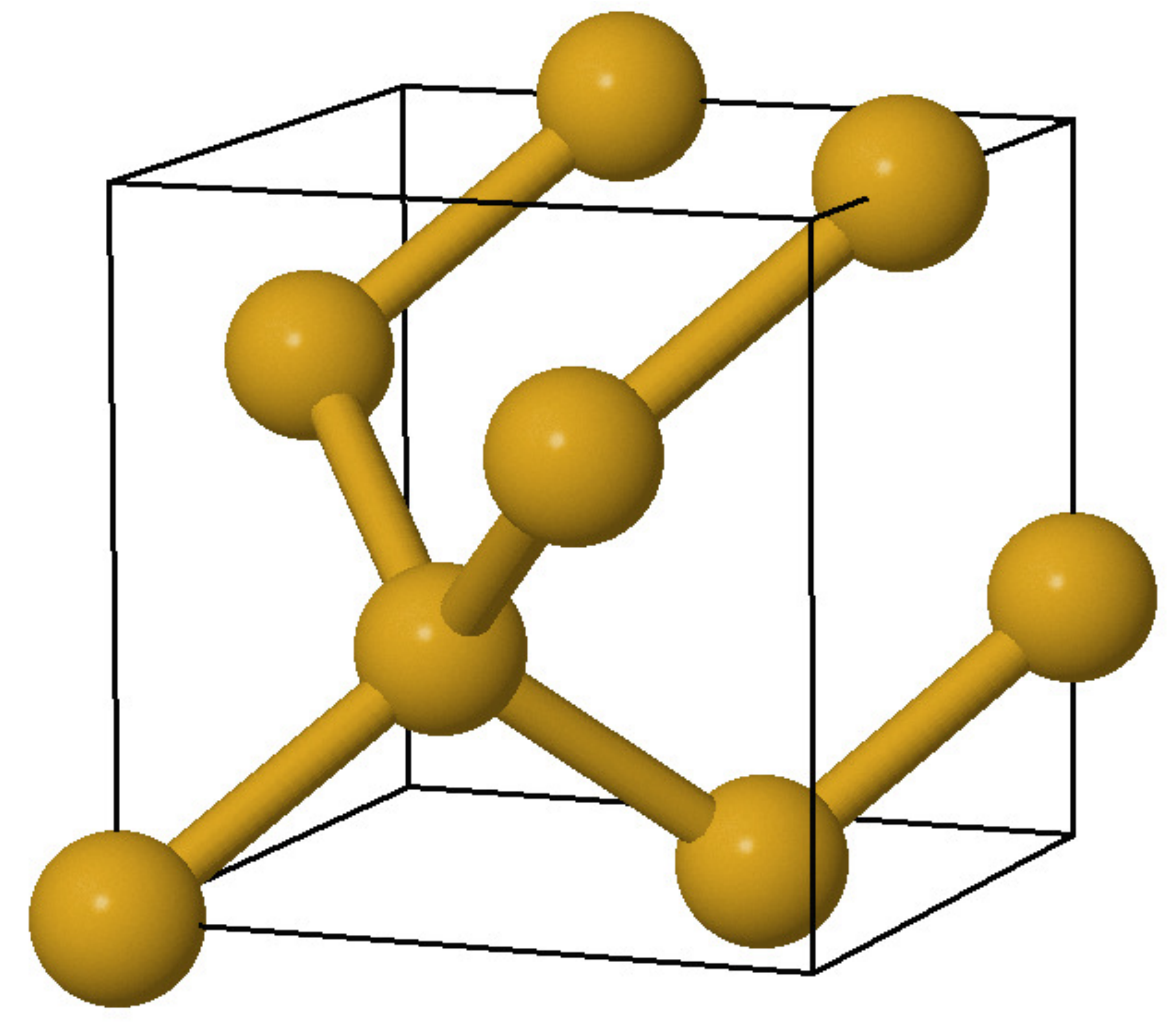} \\\hline
      & & & & & &\\   
   Graphene2D & 2D, Semimetallic & $180$ & C & $360$ & $20$ &\includegraphics[scale=0.07]{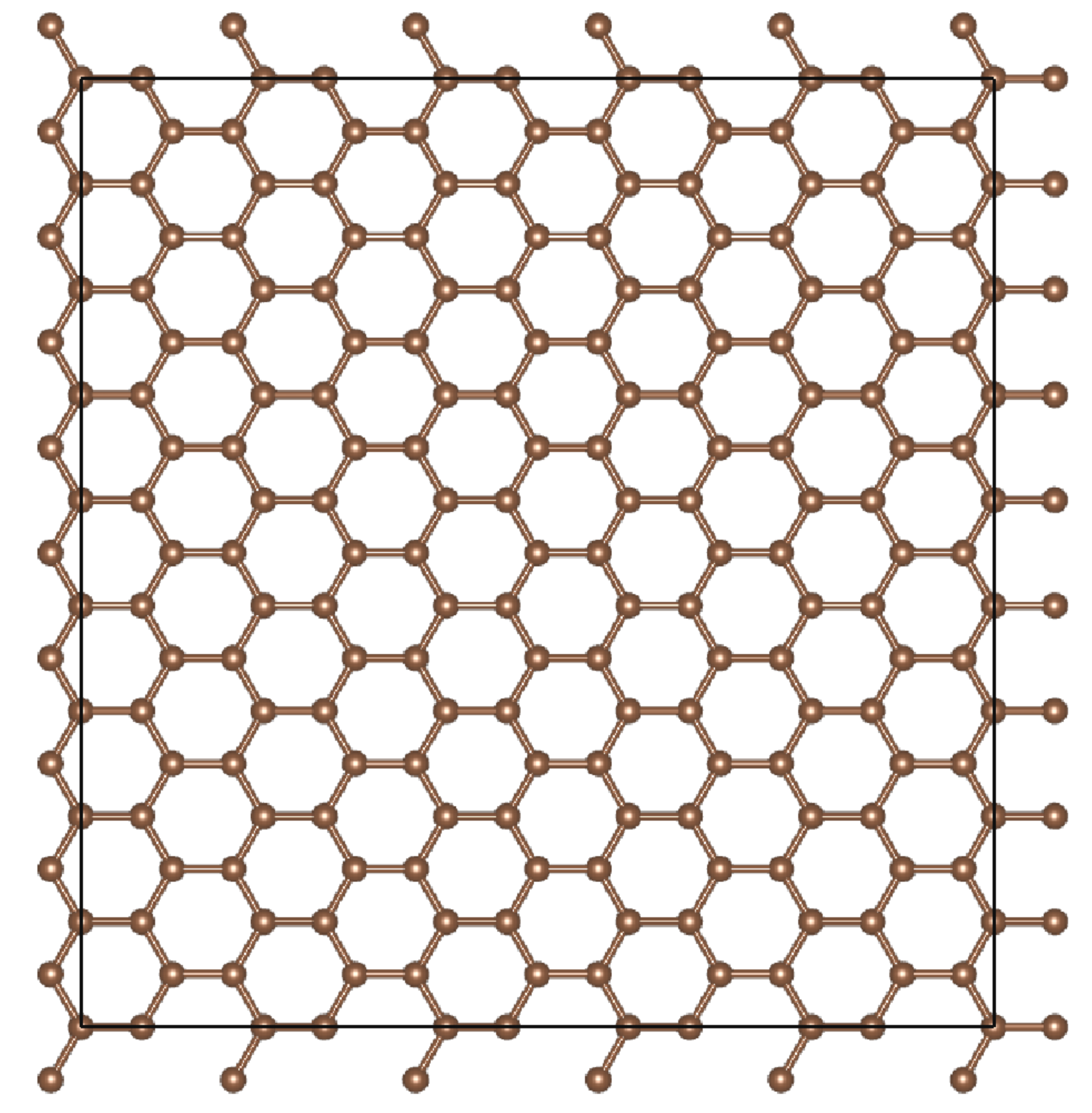} \\\hline
      & & & & & &\\
   LiBCC3D & Bulk, Metallic & $16$ & Li & $48$ & $35$  &\includegraphics[scale=0.2]{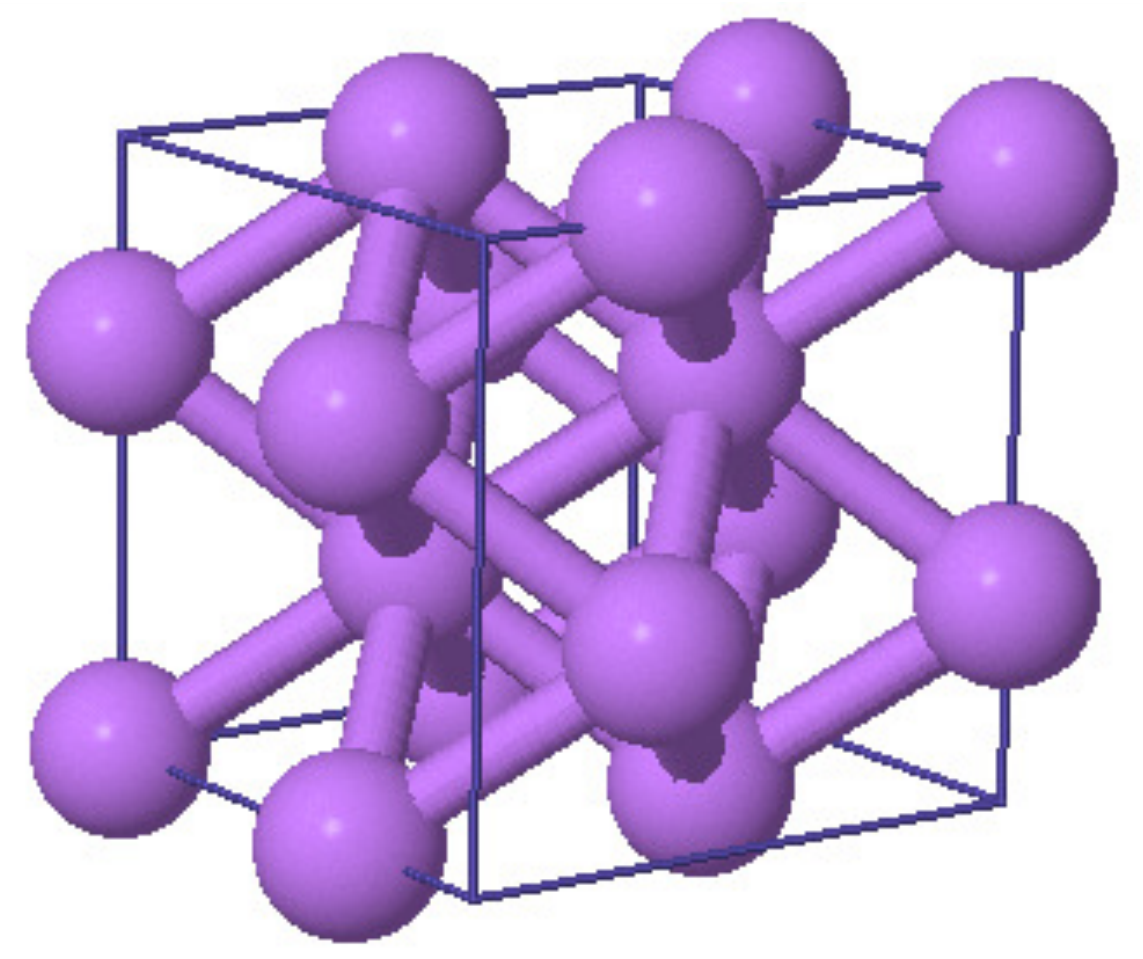}\\\hline
      & & & & & &\\
    CuFCC3D & Bulk, Metallic & $4$ & Cu & $44$ & $50$  &\includegraphics[scale=0.25]{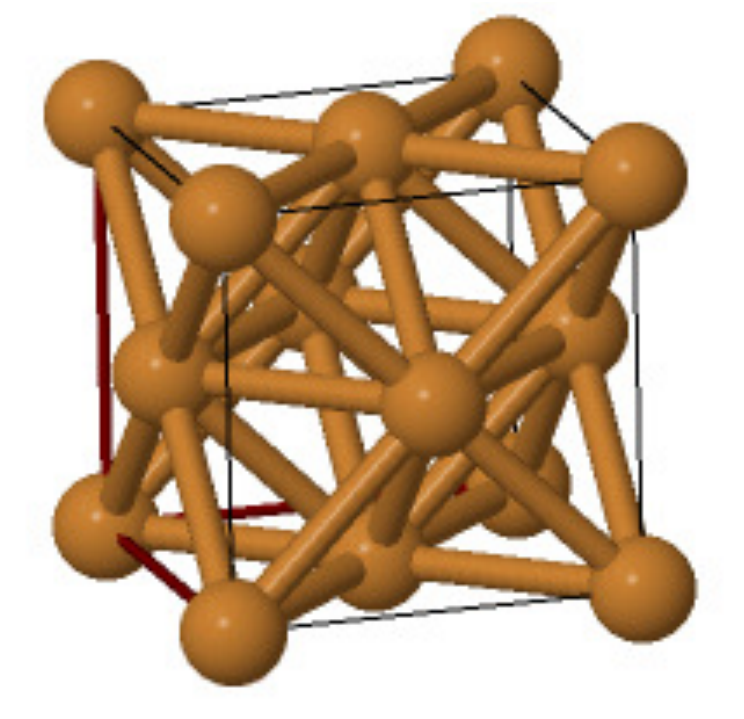} \\\hline
    \end{tabularx}
}
\end{center}
\caption{\footnotesize{Unit cells of the systems considered in this work. The simulation results presented in Section \ref{sec:Results} use supercells constructed by replicating these cells along coordinate axes to produce larger cells.}}
    \label{table:systems}
\end{table}

In order to work with larger system sizes, we have employed multiple unit cells of the aforementioned five systems replicated along the coordinate axes. Thus, $\textrm{Electrolyte3D}_{1 \times 2 \times 2}$, for example, refers to a system in which the $318$-atom unit cell has been replicated along Y and Z directions to produce a $1272$-atom bulk system; and similarly, $\textrm{Graphene2D}_{2 \times 2}$ refers to a graphene sheet containing $720$ atoms.

We have used the local density approximation for the
exchange-correlation functional with a rational function parametrization \citep{GTH_pseudoptential} of the correlation energy calculated by Ceperley-Alder\citep{ceperley_alder}. 
Hartwigsen-Goedecker-Hutter \citep{GTH_pseudoptential, GTH_relativistic} and Optimized
Norm-Conserving Vanderbilt (ONCV) pseudopotentials
\citep{hamann2013optimized, schlipf2015optimization} are employed to
remove inert core electrons from the computations. Whenever required,
SCF convergence was accelerated by means of Pulay's scheme
\citep{pulay_mixing, periodic_pulay} and an electronic temperature of
$300$ K was used for Fermi-Dirac occupation. Additionally, a Kerker
preconditioner \citep{Kresse_abinitio_MD, lin2013elliptic} was employed
to minimize charge sloshing while treating metallic systems. The
various discretization related parameters in DGDFT (specifically, the
number of ALBs per atom, DG penalty parameter, and 
fineness of real-space grid) were chosen such that chemical accuracy
could be attained \citep{lin2012adaptive, zhang2017adaptive} (i.e., error in total energies and 
forces less than $10^{-3}$ Ha/atom and $10^{-3}$ Ha/Bohr, 
respectively, relative to reference planewave results). 
This ensures that the calculations presented here are carried out 
at accuracies typical in practice.

We have typically employed $5\%$ of extra states to accommodate
fractional occupations (i.e., $N_x = 5\%$ of $N_e / 2$, or equivalently,
$N_s = 1.05 \times N_e / 2$). Unless specified
otherwise,\footnote{{In practice, for the cases considered
here, these choices ensure that the states up to index $N_1$ have
occupation numbers that differ from $1.0$ by less than $10^{-10}$. In our experience, this is quite conservative, since as long as accuracies of $10^{-7}$ or better are maintained with respect to occupation numbers, total energies and forces do not change appreciably and SCF convergence is maintained.}}  the complementary subspace calculations used the topmost $10\%$ of states, i.e., $N_t = 0.1 \times N_s$. For calculations where LOBPCG was used to compute top states, $10-15$ LOBPCG iterations per SCF step were used. For calculations using the CS2CF strategy, the order of the inner Chebyshev filter and number of inner CheFSI cycles were both set to 4. {Other parameters relevant to the CS2CF strategy were chosen according to the values shown in Table \ref{tab:parameters}.}

All calculations described here were performed on the Edison platform at the National Energy Research Scientific Computing (NERSC) center.
Edison has 5462 Cray XC30 nodes. Each node has 64 GB of memory and 24 cores partitioned
among two Intel Ivy Bridge processors, running at 2.4GHz. Edison employs a Cray Aries high-speed interconnect with Dragonfly topology for inter-node communication.

\subsection{SCF convergence and accuracy}
\label{subsec:SCF_Convergence_and_accuracy}
\begin{figure}
\subfloat[\scriptsize{$\textrm{Electrolyte3D}_{1 \times 2 \times 2}$ system ($1272$ atoms).} \label{subfig:Electrolyte_Conv}]{\scalebox{0.9}
{\begin{tikzpicture} 
\begin{axis}[
width=0.48\textwidth,
xlabel={SCF Iteration Number},
ylabel={$\textrm{Log}_{10}$ of SCF residual},
legend style={at={(0.01,0.15)},anchor=west,font=\sffamily\tiny,draw=none,fill=none,cells={anchor=west},row sep=0pt},
y label style={yshift=-10pt},
x label style={xshift=-10pt},
label style={font=\sffamily\footnotesize},
tick label style={font=\sffamily\footnotesize},
xmin=1, extra x ticks={1},
ymin=-6.5,ymax=0.5,
]
\addplot[thick, blue, mark=*] table[x index=0,y index=1]{./convergence_E318_Cheby_Ref.txt};
\addplot[thick, red, mark=o] table[x index=0,y index=1]{./convergence_E318_CS.txt};
%\addplot[thick,mark=,blue] table[x index=0,y index=#3]{\datfile{#2}};
\legend{Standard CheFSI (reference), CS Strategy};
\end{axis}
\end{tikzpicture}} 
}$\;$
\subfloat[\scriptsize{$\textrm{SiDiamond3D}_{5 \times 5 \times 5}$ system ($1000$ atoms).}\label{subfig:Silicon_Conv}]{\scalebox{0.9}
{\begin{tikzpicture} 
\begin{axis}[
width=0.48\textwidth,
xlabel={SCF Iteration Number},
ylabel={$\textrm{Log}_{10}$ of SCF residual},
legend style={at={(0.01,0.15)},anchor=west, font=\sffamily\tiny,draw=none,fill=none,cells={anchor=west},row sep=0pt},
y label style={yshift=-10pt},
x label style={xshift=-10pt},
label style={font=\sffamily\footnotesize},
tick label style={font=\sffamily\footnotesize},
xmin=1, extra x ticks={1},
ymin=-6.5,ymax=0.5,
]
\addplot[thick, blue, mark=*] table[x index=0,y index=1]{./convergence_Si_Cheby_Ref.txt};
\addplot[thick, black, mark=square] table[x index=0,y index=1]{./convergence_Si_CS_LOBPCG.txt};
\addplot[thick, red, mark=o] table[x index=0,y index=1]{./convergence_Si_CS_CheFSI.txt};
\legend{Standard CheFSI (reference), CS Strategy with LOBPCG, CS Strategy with CheFSI (CS2CF)};
\end{axis}
\end{tikzpicture}}
}\\
\subfloat[\scriptsize{$\textrm{Graphene2D}_{2 \times 2}$ system ($720$ atoms).}\label{subfig:Graphene_Conv}]{\scalebox{0.9}
{\begin{tikzpicture} 
\begin{axis}[
width=0.48\textwidth,
xlabel={SCF Iteration Number},
ylabel={$\textrm{Log}_{10}$ of SCF residual},
legend style={at={(0.001,0.15)},anchor=west, font=\sffamily\tiny,draw=none,fill=none,cells={anchor=west},row sep=0pt},
y label style={yshift=-10pt},
x label style={xshift=-10pt},
label style={font=\sffamily\footnotesize},
tick label style={font=\sffamily\footnotesize},
xmin=1, extra x ticks={1},
ymin=-6.5,ymax=0.5,
]
\addplot[thick, blue, mark=*] table[x index=0,y index=1]{./convergence_Graphene_Cheby_Ref.txt};
\addplot[thick, black, mark=square] table[x index=0,y index=1]{./convergence_Graphene_CS_LOBPCG.txt};
\addplot[thick, red, mark=o] table[x index=0,y index=1]{./convergence_Graphene_CS_CheFSI.txt};
\addplot[thick, green, mark=o] table[x index=0,y index=1]{./convergence_Graphene_CS_CheFSI_Fewer_Top.txt};
\legend{Standard CheFSI (reference), CS Strategy with LOBPCG, CS Strategy with CheFSI (CS2CF), CS2CF with fewer extra \& top states};
\end{axis}
\end{tikzpicture}}
}$\;$
\subfloat[\scriptsize{$\textrm{CuFCC3D}_{5 \times 5 \times 5}$ system ($500$ atoms).}\label{subfig:Cu_Conv}]{\scalebox{0.9}
{\begin{tikzpicture} 
\begin{axis}[
width=0.48\textwidth,
xlabel={SCF Iteration Number},
ylabel={$\textrm{Log}_{10}$ of SCF residual},
legend style={at={(0.2,0.7)},anchor=west, font=\sffamily\tiny,draw=none,fill=none,cells={anchor=west},row sep=0pt},
y label style={yshift=-10pt},
x label style={xshift=-10pt},
label style={font=\sffamily\footnotesize},
tick label style={font=\sffamily\footnotesize},
xmin=1, extra x ticks={1},
ymin=-6.5,ymax=0.5,
]
\addplot[thick, blue, mark=*] table[x index=0,y index=1]{./convergence_Cu_Cheby_Ref.txt};
\addplot[thick, black, mark=square] table[x index=0,y index=1]{./convergence_Cu_CS_LOBPCG.txt};
\addplot[thick, red, mark=o] table[x index=0,y index=1]{./convergence_Cu_CS_CheFSI.txt};
\legend{Standard CheFSI (reference), CS Strategy with LOBPCG, CS Strategy with CheFSI (CS2CF)};
\end{axis}
\end{tikzpicture}}
}\\
\subfloat[\scriptsize{$\textrm{LiBCC3D}_{4 \times 4 \times 4}$ system ($1024$ atoms).}\label{subfig:Li_Conv}]{\scalebox{0.9}
{\begin{tikzpicture} 
\begin{axis}[
width=0.48\textwidth,
xlabel={SCF Iteration Number},
ylabel={$\textrm{Log}_{10}$ of SCF residual},
legend style={at={(0.22,0.75)},anchor=west, font=\sffamily\tiny,draw=none,fill=none,cells={anchor=west},row sep=0pt},
y label style={yshift=-10pt},
x label style={xshift=-10pt},
label style={font=\sffamily\footnotesize},
tick label style={font=\sffamily\footnotesize},
xmin=1, extra x ticks={1},
ymin=-6.5,ymax=0.5,
]
\addplot[thick, blue, mark=*] table[x index=0,y index=1]{./convergence_Li_Cheby_Ref.txt};
\addplot[thick, black, mark=square] table[x index=0,y index=1]{./convergence_Li_CS_LOBPCG.txt};
\addplot[thick, red, mark=o] table[x index=0,y index=1]{./convergence_Li_CS_CheFSI.txt};
\legend{Standard CheFSI (reference), CS Strategy with LOBPCG, CS Strategy with CheFSI (CS2CF)};
\end{axis}
\end{tikzpicture}}
}
\caption{\footnotesize{SCF convergence of the complementary subspace (CS) strategy and standard CheFSI  method for systems considered in this work. The top states of the projected Hamiltonian $\tilde{H}$ can be computed using LOBPCG as well as CheFSI (the CS2CF strategy) and results for both are shown.}}
\label{fig:SCF_Convergence}
\end{figure}
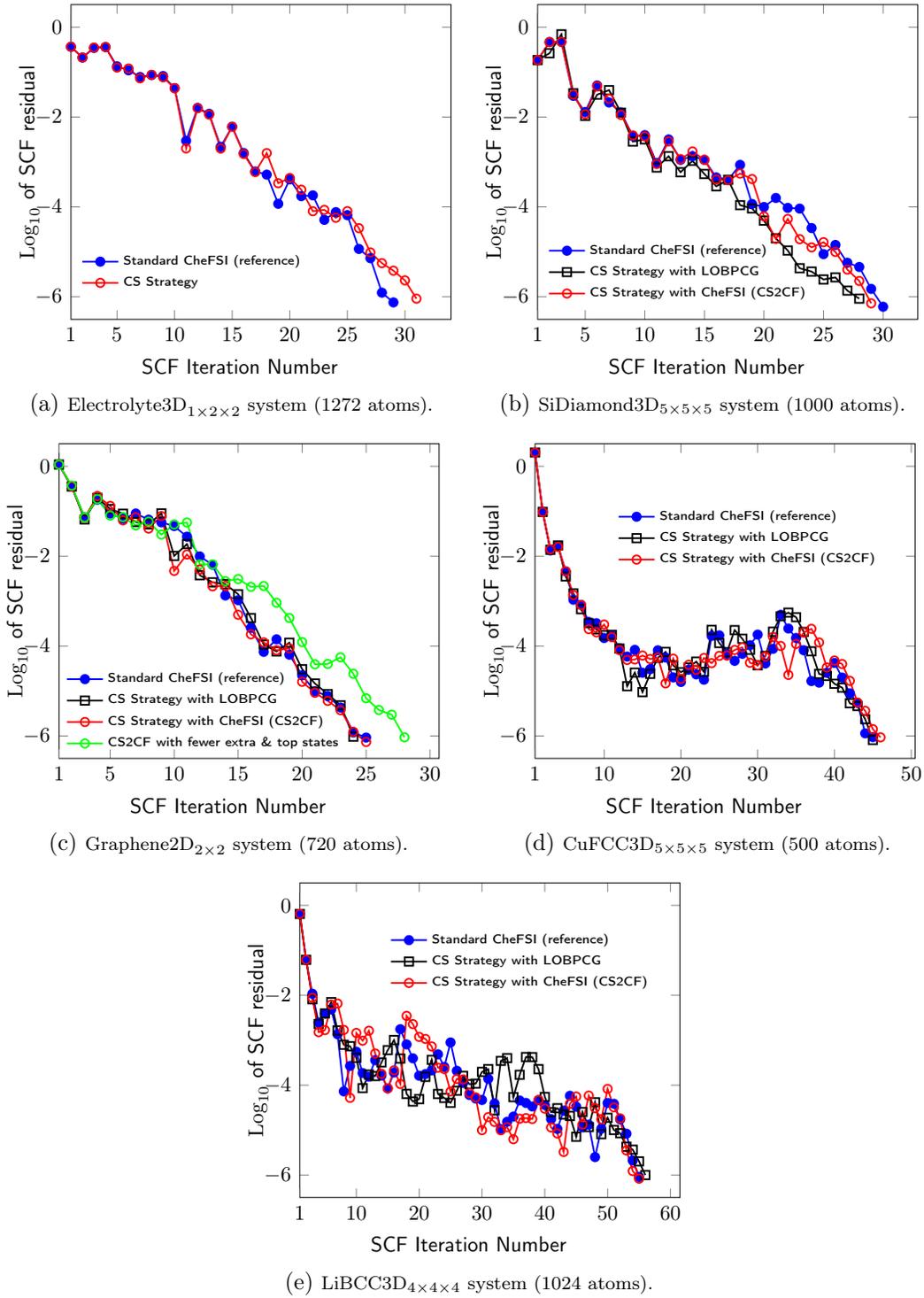

As a first test of the CS2CF methodology we first verified that it reproduces the results of the standard CheFSI methodology (with full diagonalization of the subspace Hamiltonian) with comparable SCF convergence.
Accordingly, we first compared the SCF convergence behavior of the complementary subspace strategy against the corresponding behavior of standard CheFSI (as implemented in DGDFT) for a range of systems containing from $500-1272$ atoms. Figure~\ref{fig:SCF_Convergence} shows that for all systems considered, the overall convergence behavior of the complementary subspace strategy is comparable to that of standard CheFSI, 
as should be the case since the methods are equivalent if eigenvectors are computed exactly.
Also, convergence of the complementary subspace strategy is comparable whether LOBPCG or CheFSI is used for computing the top states of the projected Hamiltonian.

Next, we verified that when SCF convergence is reached (i.e., when the relative norm of the electron density residual 
$\frac{\norm{\rho_{\text{out}} - \rho_{\text{in}}}{} }{\norm{\rho_{\text{in}}}{}} < 10^{-6}$ and 
the energy per atom has converged to $5\times 10^{-6}$ Ha), 
the energy per atom and the atomic forces obtained by the complementary subspace strategy are in agreement with the results obtained from the standard CheFSI approach 
to well below discretization error ($\sim 10^{-3}$ Ha/atom). 
Table \ref{tab:scf_accuracy} shows that for the above systems, the energy per atom difference is on the order of $10^{-5}$ Ha or less while the maximum difference in force components is on the order of $10^{-4}$ Ha/Bohr or less, an order of magnitude or more below discretization error.
\begin{table}
\begin{center}
\scriptsize{
\begin{tabular}{| c | c | c | c | c |}
  \hline
  {} & \multicolumn{2}{ c |}{{}} & \multicolumn{2}{ c |}{{}} \\ 
  {} & \multicolumn{2}{ c |}{{CS Strategy with LOBPCG}} & \multicolumn{2}{ c |}{{CS Strategy with CheFSI}} \\
   {} & \multicolumn{2}{ c |}{{for top states}} & \multicolumn{2}{ c |}{{for top states (CS2CF strategy)}} \\
  {{System}} & \multicolumn{2}{ c |}{{}} & \multicolumn{2}{ c |}{{}} \\\cline{2-5}
   & & & & \\
  & Energy per atom & Max. force component & Energy per atom & Max. force component \\
  & difference (Ha) & difference (Ha/Bohr) &  difference (Ha) & difference (Ha/Bohr)  \\
   & & & & \\\cline{1-5}
 & & & & \\
  $\textrm{Electrolyte3D}_{1 \times 2 \times 2}$  & $5 \times 10^{-5}$ & $2 \times 10^{-4}$ & $5 \times 10^{-5}$  & $2 \times 10^{-4}$\\ & & & & \\\cline{1-5}
& & & & \\
  $\textrm{SiDiamond3D}_{5 \times 5 \times 5}$  & $4 \times 10^{-6}$ & $4 \times 10^{-5}$ & $2 \times 10^{-6}$ & $1 \times 10^{-5}$\\ & & & & \\\cline{1-5}
& & & & \\
  $\textrm{Graphene2D}_{1 \times 2 \times 2}$  & $7 \times 10^{-6}$ & $8 \times 10^{-5}$ & $9 \times 10^{-6}$ & $8 \times 10^{-5}$\\ & & & & \\\cline{1-5}
  & & & & \\
 $\textrm{CuFCC3D}_{5 \times 5 \times 5}$  &$8 \times 10^{-6}$ & $9 \times 10^{-5}$ & $4 \times 10^{-6}$ & $7 \times 10^{-5}$\\ & & & & \\\cline{1-5}
  & & & & \\
  $\textrm{LiBCC3D}_{4 \times 4 \times 4}$   & $7 \times 10^{-6}$ & $9 \times 10^{-5}$ & $5 \times 10^{-6}$ & $8 \times 10^{-5}$\\ & & & & \\
  \hline
\end{tabular}
}
\end{center}
\caption{\footnotesize{Accuracy of complementary subspace (CS) strategy 
using LOBPCG and CheFSI (CS2CF) methods to compute top states. 
Shown are energy per atom and force component differences from
standard CheFSI results. 
Differences are well below discretization error in all cases. 
}}
\label{tab:scf_accuracy}
\end{table}

It is worth pointing out a difference between the scenario depicted in Fig.~\ref{subfig:Electrolyte_Conv} and the other cases shown in Fig.~\ref{fig:SCF_Convergence}. As discussed in 
Section~\ref{subsubsec:DM_and_Proj_DM}, it is not necessary to carry out the subspace diagonalization step or its complementary subspace counterpart while treating insulating systems (like the electrolyte considered in Fig.~\ref{subfig:Electrolyte_Conv}). Hence, there is no distinction between the LOBPCG and CheFSI based complementary subspace strategies in the case of Fig.~\ref{subfig:Electrolyte_Conv}, thus leading to the single curve for the complementary subspace strategy.

We note also the case of graphene shown in Fig.~\ref{subfig:Graphene_Conv}. 
Semimetallic systems such as this tend to have relatively few fractionally occupied states 
near the Fermi level at moderate electronic temperature. \citep{wiki_semimetal} Hence, it is possible to apply (finite electronic temperature) smearing to such systems with fewer extra states in the calculation, thus reducing $N_s$. Furthermore, the complementary subspace strategy can be made to use fewer top states, thus reducing $N_t$. This reduces computational effort without significantly impacting accuracy or SCF convergence. 
As Fig.~\ref{subfig:Graphene_Conv} shows, even with fewer extra and top states in the computation (specifically, the CS2CF strategy used $N_s = 1.025 \times N_e/2$ rather than the usual $N_s = 1.05 \times N_e/2$,  and $N_t = 5\%$ of $N_s$), SCF convergence is not significantly affected. We also verified that 
the converged energies and forces agreed with reference CheFSI results to well below discretization error. 
As demonstrated in the next section, however, the computational gains from using fewer states can be quite noticeable, especially for large systems.
\subsection{Computational efficiency and parallel scaling}
\label{subsec:efficiency_and_scaling}
We now carry out a systematic comparison of the computational efficiency and parallel scaling of the 
new CS2CF and standard CheFSI methodologies. 

In order to carry out the comparison, we investigate the wall time for the construction and 
solution of the subspace problem for a number of large systems. Within the context of the CheFSI or CS2CF strategies, this is the time spent in the sequence of computational steps that lead to the diagonal blocks of the (full) density matrix, after the Chebyshev polynomial filtered vectors given by the columns of $Y$  have been computed using $H$. Therefore, this time includes contributions from steps that are common to both standard CheFSI and CS2CF strategies, such as orthonormalization of the Chebyshev polynomial filtered vectors (i.e., columns of $Y$) and formation of the projected Hamiltonian $\tilde{H} = Y^T (HY)$ from the vector blocks $Y$ and $Z = HY$. Additionally, for the standard CheFSI method, it includes the time spent on  subspace diagonalization and subspace rotation steps. In contrast, for the CSF2CF strategy, it includes the time spent on computing the top states of the projected Hamiltonian (using the inner level of CheFSI on $\tilde{H}$), and any additional computation required for evaluating the diagonal blocks of the density matrix (see Eq.~\ref{eq:P_CS_expression}). Thus, within the context of the standard CheFSI or CS2CF strategies, the subspace problem construction and solution wall time provides an estimate of the total wall time spent on every SCF step in (distributed) dense linear algebra operations.

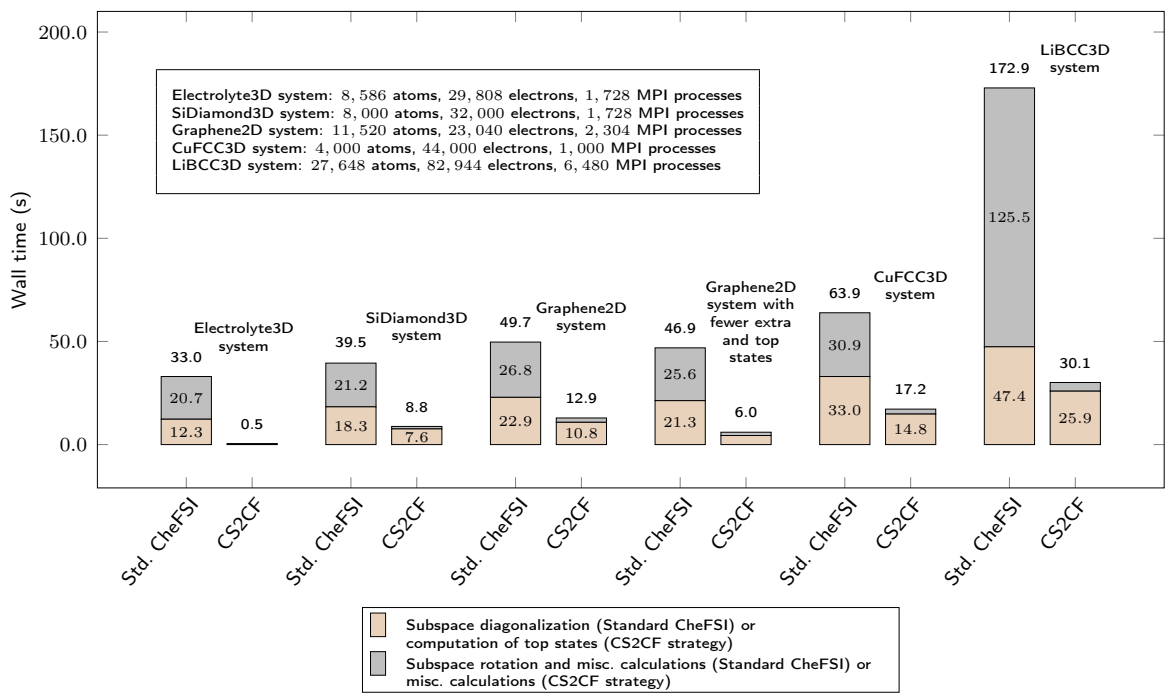
\begin{figure}
\subfloat{\scalebox{0.95}
{\begin{tikzpicture} 
\pgfkeys{
    /pgf/number format/precision=1, 
    /pgf/number format/fixed zerofill=true
}
\begin{axis}[ybar stacked,font=\sffamily, 
width=\textwidth,
height=0.5\textwidth,
%xmajorticks=false,
bar width=20pt,
ymax = 210,
nodes near coords,
%every text node part/.style={align=center},
every node near coord/.style={
      check for zero/.code={
        \pgfkeys{/pgf/fpu=true}
        \pgfmathparse{\pgfplotspointmeta-4.5}
        \pgfmathfloatifflags{\pgfmathresult}{-}{
           \pgfkeys{/tikz/coordinate}
        }{}
        \pgfkeys{/pgf/fpu=false}
      }, check for zero, font=\tiny},
%every node near coord/.append style={font=\tiny},
y label style={yshift=-5pt},
ylabel={Wall time (s)},
xtick=data,
xticklabels={Std.~CheFSI, CS2CF,  Std.~CheFSI, CS2CF, Std.~CheFSI, CS2CF, Std.~CheFSI, CS2CF, Std.~CheFSI, CS2CF, Std.~CheFSI, CS2CF},
label style={font=\sffamily\scriptsize},
tick label style={font=\sffamily\scriptsize},
x tick label style={rotate=50,anchor=east, yshift= -5.5, xshift = 2.5},
legend style={at={(0.5,-0.25)},  anchor=north,legend columns=1,
fill=none,font=\sffamily, cells={anchor=west},row sep=1pt}] 
% Plot data
\addplot [fill=brown!35,] coordinates {(10, 12.3) (50,0) (110, 18.3) (150, 7.6) (210, 22.9) (250, 10.8) (310, 21.3) (350, 4.4) (410, 33.0) (450,14.8) (510, 47.4) (550, 25.9)}; 
\addplot [fill=black!25,] coordinates  {(10, 20.7) (50,0.5) (110, 21.2) (150,1.2) (210, 26.8) (250, 2.1) (310, 25.6) (350, 1.6) (410, 30.9) (450,2.4) (510, 125.5) (550,4.2)};

% % Print the totals
\draw (10,50) node [font=\sffamily, draw = none, align=left,   below] {\tiny{33.0}};
\draw (50, 17.5) node [font=\sffamily, draw = none, align=left,   below] {\tiny{0.5}};

\draw (110,57) node [font=\sffamily, draw = none, align=left,   below] {\tiny{39.5}};
\draw (150, 25.5) node [font=\sffamily, draw = none, align=left,   below] {\tiny{8.8}};

\draw (210,66.7) node [font=\sffamily, draw = none, align=left,   below] {\tiny{49.7}};
\draw (250, 29.9) node [font=\sffamily, draw = none, align=left,   below] {\tiny{12.9}};

\draw (310, 63.9) node [font=\sffamily, draw = none, align=left,   below] {\tiny{46.9}};
\draw (350, 23.0) node [font=\sffamily, draw = none, align=left,   below] {\tiny{6.0}};

\draw (410, 80.9) node [font=\sffamily, draw = none, align=left,   below] {\tiny{63.9}};
\draw (450, 34.2) node [font=\sffamily, draw = none, align=left,   below] {\tiny{17.2}};

\draw (510,189.9) node [font=\sffamily, draw = none, align=left,   below] {\tiny{172.9}};
\draw (550, 47.1) node [font=\sffamily, draw = none, align=left,   below] {\tiny{30.1}};

%% Print some labels
\draw (45 ,65.0) node [font=\sffamily, align=left,   below] {\tiny{\begin{tabular}{c} {Electrolyte3D}\\ system \end{tabular}}};

\draw (150 ,70.0) node [font=\sffamily, align=left,   below] {\tiny{\begin{tabular}{c} {SiDiamond3D}\\ system \end{tabular}}};

\draw (250 , 75.0) node [font=\sffamily, align=left,   below] {\tiny{\begin{tabular}{c} {Graphene2D}\\ system \end{tabular}}};

\draw (350 + 3, 85.0) node [font=\sffamily, align=left,   below] {\tiny{\begin{tabular}{c} {Graphene2D}\\ system with \\ fewer extra \\ and  top \\ states\end{tabular}}};

\draw (450 , 90.0) node [font=\sffamily, align=left,   below] {\tiny{\begin{tabular}{c} {CuFCC3D}\\ system \end{tabular}}};

\draw (550 , 200.0) node [font=\sffamily, align=left,   below] {\tiny{\begin{tabular}{c} {LiBCC3D}\\ system \end{tabular}}};

% % Legend
\legend{\tiny{\begin{tabular}{l}Subspace diagonalization (Standard CheFSI) or\\ computation of top states (CS2CF strategy)\end{tabular}}, \tiny{\begin{tabular}{l}Subspace rotation and misc.~calculations (Standard CheFSI) or\\ misc.~calculations (CS2CF strategy) \end{tabular}}}

% % Insert a table here
\draw (175 , 187.0) node [font=\sffamily, align=left,   below] {\tiny{\begin{tabular}{| l |}
\hline \\
{Electrolyte3D system:} {$8,586$ atoms, $29,808$ electrons, $1,728$ MPI processes}\\
{SiDiamond3D system:} {$8,000$ atoms, $32,000$ electrons, $1,728$ MPI processes}\\
{Graphene2D system:} {$11,520$ atoms, $23,040$ electrons, $2,304$ MPI processes}\\
{CuFCC3D system:} {$4,000$ atoms, $44,000$ electrons, $1,000$ MPI processes}\\
{LiBCC3D system:} {$27,648$ atoms, $82,944$ electrons, $6,480$ MPI processes} \\
\\ \hline
\end{tabular}}};
\end{axis} 
 \end{tikzpicture}}
}
\caption{\footnotesize{Wall times associated with solution of the subspace problem using standard CheFSI (subspace diagonalization, subspace rotation, miscellaneous calculations) and new CS2CF (computation of top states and miscellaneous calculations) strategies for a few large systems. }}
\label{fig:Block_Chart_1}
\end{figure}

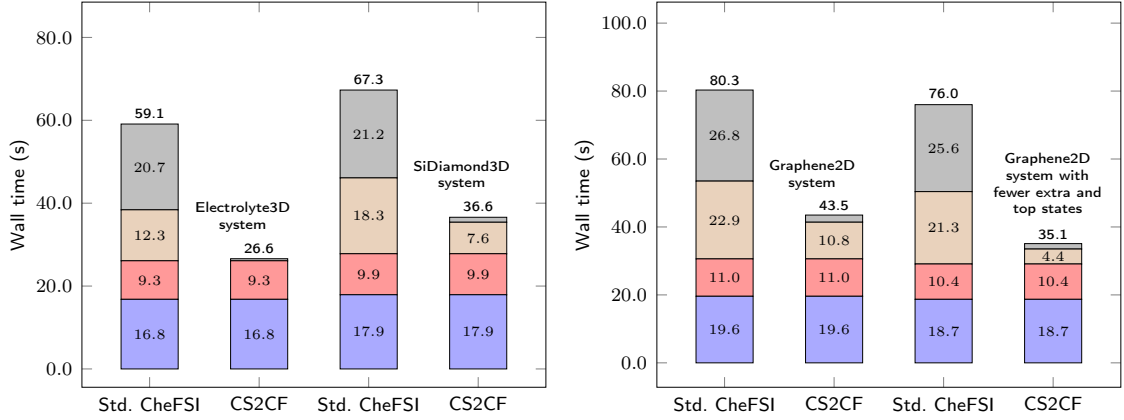
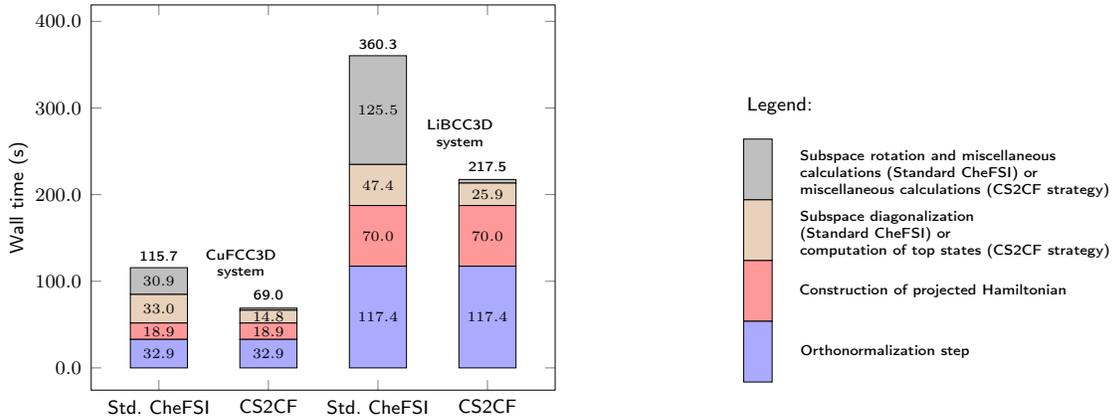
\begin{figure}
\subfloat[\scriptsize{$\textrm{Electrolyte3D}_{3 \times 3 \times 3}$ system ($8,586$ atoms, $29,808$ electrons, left two plots) and $\textrm{SiDiamond3D}_{10 \times 10 \times 10}$ system  ($8,000$ atoms, $32,000$ electrons, right two plots). $1,728$ MPI processes used in all cases.}\label{subfig:Electrolyte_Block_Chart}]{\scalebox{0.90}
{\begin{tikzpicture} 
\pgfkeys{
    /pgf/number format/precision=1, 
    /pgf/number format/fixed zerofill=true
}
\begin{axis}[ybar stacked,font=\sffamily, 
%xmajorticks=false,
bar width=24pt,
nodes near coords,
%every text node part/.style={align=center},
every node near coord/.style={
      check for zero/.code={
        \pgfkeys{/pgf/fpu=true}
        \pgfmathparse{\pgfplotspointmeta-3}
        \pgfmathfloatifflags{\pgfmathresult}{-}{
           \pgfkeys{/tikz/coordinate}
        }{}
        \pgfkeys{/pgf/fpu=false}
      }, check for zero, font=\tiny},
%every node near coord/.append style={font=\tiny},
y label style={yshift=-5pt},
ylabel={Wall time (s)},
xtick=data,
xticklabels={Std.~CheFSI, CS2CF, Std.~CheFSI, CS2CF},
label style={font=\sffamily\scriptsize},
tick label style={font=\sffamily\scriptsize},
%x tick label style={rotate=0,anchor=east},
enlarge x limits={abs=1cm},
enlarge y limits={abs=1.3cm},
legend style={at={(0.5,-0.10)}, anchor=north,legend columns=1,
draw=none,fill=none,font=\sffamily, cells={anchor=west},row sep=0pt}] 
% Plot data
\addplot [fill={rgb:blue,2;white,4}]coordinates {(1,16.8) (2,16.8) (3,17.9) (4, 17.9) }; 
\addplot [fill={rgb:red,2;white,3}] coordinates {(1, 9.3) (2,9.3) (3, 9.9) (4, 9.9) }; 
\addplot [fill=brown!35,] coordinates {(1, 12.3) (2,0) (3, 18.3) (4, 7.6) }; 
\addplot [fill=black!25,] coordinates  {(1, 20.7) (2,0.5) (3, 21.2) (4,1.2)}; 
% Print the totals
\draw (1,59.1+6) node [font=\sffamily, draw = none, align=left,   below] {\tiny{59.1}};
\draw (2,26.6+6) node [font=\sffamily, draw = none, align=left,   below] {\tiny{26.6}};
\draw (3,67.3+6) node [font=\sffamily, draw = none, align=left,   below] {\tiny{67.3}};
\draw (4,36.6+6) node [font=\sffamily, draw = none, align=left,   below] {\tiny{36.6}};
% Print some labels
\draw (1.85 ,43.0) node [font=\sffamily, align=left,   below] {\tiny{\begin{tabular}{c} {Electrolyte3D}\\ system \end{tabular}}};
\draw (3.85 ,53.0) node [font=\sffamily, align=left,   below] {\tiny{\begin{tabular}{c} {SiDiamond3D}\\ system \end{tabular}}};
%\legend{ $\;$ Orthonormalization step, $\;$ Construction of projected Hamiltonian,  $\;$ Direct solution of subspace problem (Standard CheFSI) or Computation of top states (CS2CF strategy),  $\;$ Subspace rotation and misc.~calculations (Standard CheFSI) or misc.~calculations (CS2CF strategy)}
\end{axis} 
 \end{tikzpicture}}}$\;$
\subfloat[\scriptsize{$\textrm{Graphene2D}_{8 \times 8}$ system ($11,520$ atoms, $23,040$ electrons). Right two figures use fewer extra and top states. $2,304$ MPI processes used in all cases.}\label{subfig:Graphene_Block_Chart}]{\scalebox{0.90}
{\begin{tikzpicture} 
\pgfkeys{
    /pgf/number format/precision=1, 
    /pgf/number format/fixed zerofill=true
}
\begin{axis}[ybar stacked,font=\sffamily, 
%xmajorticks=false,
bar width=24pt,
nodes near coords,
%every text node part/.style={align=center},
every node near coord/.style={
      check for zero/.code={
        \pgfkeys{/pgf/fpu=true}
        \pgfmathparse{\pgfplotspointmeta-3}
        \pgfmathfloatifflags{\pgfmathresult}{-}{
           \pgfkeys{/tikz/coordinate}
        }{}
        \pgfkeys{/pgf/fpu=false}
      }, check for zero, font=\tiny},
%every node near coord/.append style={font=\tiny},
y label style={yshift=-5pt},
ylabel={Wall time (s)},
xtick=data,
xticklabels={Std.~CheFSI, CS2CF, Std.~CheFSI, CS2CF},
label style={font=\sffamily\scriptsize},
tick label style={font=\sffamily\scriptsize},
%x tick label style={rotate=0,anchor=east},
enlarge x limits={abs=1cm},
enlarge y limits={abs=1.3cm},
legend style={at={(0.5,-0.10)}, anchor=north,legend columns=1,
draw=none,fill=none,cells={anchor=west},row sep=0pt}] 
% Plot data
\addplot [fill={rgb:blue,2;white,4}] coordinates {(1,19.6) (2,19.6) (3,18.7) (4, 18.7) }; 
\addplot [fill={rgb:red,2;white,3}]  coordinates {(1, 11.0) (2, 11.0) (3, 10.4) (4, 10.4) }; 
\addplot [fill=brown!35,]  coordinates {(1, 22.9) (2, 10.8) (3, 21.3) (4, 4.4) }; 
\addplot [fill=black!25,]  coordinates {(1, 26.8) (2, 2.1) (3, 25.6) (4,1.6)}; 
% Print the totals
\draw (1,80.3+7) node [font=\sffamily, draw = none, align=left,   below] {\tiny{80.3}};
\draw (2,43.5+7) node [font=\sffamily, draw = none, align=left,   below] {\tiny{43.5}};
\draw (3,76.0+7.5) node [font=\sffamily, draw = none, align=left,   below] {\tiny{76.0}};
\draw (4,35.1+7) node [font=\sffamily, draw = none, align=left,   below] {\tiny{35.1}};
% Print some labels
\draw (1.80 ,63.0) node [font=\sffamily, align=left,   below] {\tiny{\begin{tabular}{c} {Graphene2D}\\ system \end{tabular}}};
\draw (3.95 ,65.0) node [font=\sffamily, align=left,   below] {\tiny{\begin{tabular}{c} {Graphene2D}\\ system with \\ fewer extra and \\ top states\end{tabular}}};
%\legend{ $\;$ Orthonormalization step, $\;$ Construction of projected Hamiltonian,  $\;$ Direct solution of subspace problem (Standard CheFSI) or Computation of top states (CS2CF strategy),  $\;$ Subspace rotation and misc.~calculations (Standard CheFSI) or misc.~calculations (CS2CF strategy)}
\end{axis} 
\end{tikzpicture}}}\\
\subfloat[\scriptsize{$\textrm{CuFCC3D}_{10 \times 10 \times 10}$ system ($4000$ atoms, $44,000$ electrons, $1,000$ MPI processes, left two plots) and $\textrm{LiBCC3D}_{12 \times 12 \times 12}$ system  ($27,648$ atoms, $82,944$ electrons, $6,480$ MPI processes, right two plots). }\label{subfig:Metals_Block_Chart}]{\scalebox{0.90}
{\begin{tikzpicture} 
\pgfkeys{
    /pgf/number format/precision=1, 
    /pgf/number format/fixed zerofill=true
}
\begin{axis}[ybar stacked,font=\sffamily, 
%xmajorticks=false,
bar width=24pt,
nodes near coords,
%every text node part/.style={align=center},
every node near coord/.style={
      check for zero/.code={
        \pgfkeys{/pgf/fpu=true}
        \pgfmathparse{\pgfplotspointmeta-5}
        \pgfmathfloatifflags{\pgfmathresult}{-}{
           \pgfkeys{/tikz/coordinate}
        }{}
        \pgfkeys{/pgf/fpu=false}
      }, check for zero, font=\tiny},
%every node near coord/.append style={font=\tiny},
y label style={yshift=-5pt},
ylabel={Wall time (s)},
xtick=data,
xticklabels={Std.~CheFSI, CS2CF, Std.~CheFSI, CS2CF},
label style={font=\sffamily\scriptsize},
tick label style={font=\sffamily\scriptsize},
%x tick label style={rotate=0,anchor=east},
enlarge x limits={abs=1cm},
enlarge y limits={abs=0.75cm},
legend style={at={(1.7,0.8)}, anchor=north,legend columns=1,
draw=none,fill=none,cells={align = left, anchor=west},row sep=0pt}] 
% Plot data
\addplot [fill={rgb:blue,2;white,4}] coordinates {(1,32.9) (2,32.9) (3,117.4) (4, 117.4) }; 
\addplot [fill={rgb:red,2;white,3}]  coordinates {(1, 18.9) (2,18.9) (3, 70.0) (4, 70.0) }; 
\addplot [fill=brown!35,] coordinates {(1, 33.0) (2,14.8) (3, 47.4) (4, 25.9) }; 
\addplot [fill=black!25,] coordinates {(1, 30.9) (2,2.4) (3, 125.5) (4,4.2)}; 
% Print the totals
\draw (1,115.7+30) node [font=\sffamily, draw = none, align=left,   below] {\tiny{115.7}};
\draw (2,69.0+32) node [font=\sffamily, draw = none, align=left,   below] {\tiny{69.0}};
\draw (3,360.3+30) node [font=\sffamily, draw = none, align=left,   below] {\tiny{360.3}};
\draw (4,217.5+32) node [font=\sffamily, draw = none, align=left,   below] {\tiny{217.5}};
% Print some labels
\draw (1.75 ,150.0) node [font=\sffamily, align=left,   below] {\tiny{\begin{tabular}{c} {{CuFCC3D}}\\ system \end{tabular}}};
\draw (3.75 ,300) node [font=\sffamily, align=left,   below] {\tiny{\begin{tabular}{c} {LiBCC3D}\\ system \end{tabular}}};
%\legend{ $\;$ Orthonormalization step, $\;$ Construction of projected Hamiltonian,  $\;$ Direct solution of subspace problem (Standard CheFSI) \\ or Computation of top states (CS2CF strategy),  $\;$ Subspace rotation and misc.~calculations (Standard CheFSI) \\ or misc.~calculations (CS2CF strategy)}
\end{axis} 
\end{tikzpicture}}}$\quad\quad\quad\quad$
\subfloat{\scalebox{0.90}
{\begin{tikzpicture} 
\begin{axis}[ybar stacked,
%xmajorticks=false,
bar width=12pt,
nodes near coords,
ticks=none,
axis line style={draw=none},
xmax = 2.5,
%every text node part/.style={align=center},
every node near coord/.style={
      check for zero/.code={
        \pgfkeys{/pgf/fpu=true}
        \pgfmathparse{\pgfplotspointmeta-5}
        \pgfmathfloatifflags{\pgfmathresult}{-}{
           \pgfkeys{/tikz/coordinate}
        }{}
        \pgfkeys{/pgf/fpu=false}
      }, check for zero, font=\tiny},
%every node near coord/.append style={font=\tiny},
y label style={yshift=-5pt},
ylabel={},
%xtick=data,
%xticklabels={Std.~CheFSI, CS2CF, Std.~CheFSI, CS2CF},
%x tick label style={rotate=0,anchor=east},
enlarge x limits={abs=1cm},
enlarge y limits={abs=1.5cm},
legend style={at={(1.7,0.8)}, anchor=north,legend columns=1,
draw=none,fill=none,cells={align = left, anchor=west},row sep=0pt}] 
% Plot data
\addplot [fill={rgb:blue,2;white,4}] coordinates {(1, 1) }; 
\addplot [fill={rgb:red,2;white,3}]  coordinates {(1, 1) }; 
\addplot [fill=brown!35,] coordinates {(1, 1) }; 
\addplot [fill=black!25,] coordinates {(1, 1) }; 
% Print some labels
\draw (1.58 ,0.8) node [font=\sffamily, align=left,   below] {\tiny{\begin{tabular}{l} Orthonormalization step \end{tabular}}};
\draw (1.8 ,1.8) node [font=\sffamily, align=left,   below] {\tiny{\begin{tabular}{l} Construction of projected Hamiltonian \end{tabular}}};
\draw (1.9 ,3.00) node [font=\sffamily, align=left,   below] {\tiny{\begin{tabular}{l} Subspace diagonalization \\(Standard CheFSI)  or \\computation of top states (CS2CF strategy)\end{tabular}}};
\draw (1.90 ,4.0) node [font=\sffamily, align=left,   below] {\tiny{\begin{tabular}{l} Subspace rotation and  miscellaneous\\ calculations (Standard CheFSI)  or \\miscellaneous calculations (CS2CF strategy)\end{tabular}}};
\draw (1.1 ,4.9) node [font=\sffamily, align=left,   below] {\scriptsize{\begin{tabular}{l} Legend: \end{tabular}}};
%\legend{ $\;$ Orthonormalization step, $\;$ Construction of projected Hamiltonian,  $\;$ Direct solution of subspace problem (Standard CheFSI) \\ or Computation of top states (CS2CF strategy),  $\;$ Subspace rotation and misc.~calculations (Standard CheFSI) \\ or misc.~calculations (CS2CF strategy)}
\end{axis} 
\end{tikzpicture}}}
\caption{\footnotesize{Wall times associated with construction and solution of the subspace problem for 
large systems using standard CheFSI and new CS2CF methods. 
Total wall times and contributions of key steps are shown.}}
\label{fig:Block_Chart_2}
\end{figure}

The systems considered for this comparison contain between $4,000$ and $27,648$ atoms and between $23,040$ and $82,944$ electrons. In each case, an identical number of computational cores were allocated to both  methods\footnote{As detailed in previous work\citep{banerjee2016chebyshev}, the number of MPI processes allocated for construction and solution of the subspace problem in DGDFT is typically set equal to the number of DG elements used in the particular calculation. The results shown in Figures~\ref{fig:Block_Chart_1} and \ref{fig:Block_Chart_2} were obtained following this practice. However, fewer or more processors can be employed as needed. Accordingly, results in the following sections often use a larger number of processors for the subspace problem to bring down wall times for larger-scale calculations.} and the ScaLAPACK process grids used were kept as close to square geometries as possible. The results are shown in Figures~\ref{fig:Block_Chart_1} and \ref{fig:Block_Chart_2}.

From Figure~\ref{fig:Block_Chart_1} we see that for the large non-insulating systems considered here, the CS2CF strategy is able to bring down the wall time for the subspace diagonalization and subspace rotation steps in the standard CheFSI method by factors of $3.7$--$7.8$. Additionally, for the particular case of the $\textrm{Electrolyte3D}_{3 \times 3 \times 3}$ insulating system, the CS2CF strategy eliminates these steps altogether, and brings down the wall time by a factor exceeding $60$. This dramatic reduction of the wall times of key steps of the standard CheFSI method leads us to expect that the overall subspace construction and solution time for these systems will be reduced significantly as well. This expectation turns out to be correct. From Figure~\ref{fig:Block_Chart_2}, we see that the overall subspace problem construction and solution wall time is brought down by a factor of $\sim 1.7$ to $2.2$. These computational wall time savings are particularly significant in light of the fact that they occur on every SCF step. From the figure, it is also evident that the overall savings due to the replacement of the subspace diagonalization and subspace rotation steps by the corresponding CS2CF steps are most significant for systems in which these steps are the largest contributors to the subspace problem wall time. For the largest system considered here, i.e.,   $\textrm{LiBCC3D}_{12 \times 12 \times 12}$ ($27,648$ atoms, $82,944$ electrons), the orthonormalization cost contributes to the subspace problem construction wall time in a significant manner and so the overall savings due to the CS2CF strategy, while substantial ($\sim$5.7x reduction in the time spent on the subspace diagonalization and subspace rotation steps, as shown in 
Fig.~\ref{fig:Block_Chart_1}), are somewhat smaller (i.e., $\sim$1.7x reduction in overall subspace problem wall time) compared to other cases in Figure \ref{fig:Block_Chart_2}. 

In light of earlier comments (see Sections \ref{subsubsec:chefsi_top_states} and  \ref{subsec:SCF_Convergence_and_accuracy}), it is worth pointing out the computational benefits of using fewer extra and top states in the CS2CF strategy. As can be seen in Figures~\ref{fig:Block_Chart_1} and \ref{subfig:Graphene_Block_Chart}, on reducing the number of extra states (i.e., $N_x$), and consequently,  the total number of states $N_s$, the variations in wall times for the steps of the standard CheFSI strategy are not particularly significant. However, lowering $N_x$ also allows us to lower the number of top states (i.e., $N_t$) in the CS2CF strategy, and this leads to a significant savings. Specifically, as Figure~\ref{fig:Block_Chart_1} shows, the total wall time spent in the routines associated with the CS2CF strategy decreases by more than a factor of $2$ due to the value of $N_t$ being halved.\footnote{Such a striking demonstration of the virtue of lowering $N_x$ and $N_t$ in the CS2CF strategy led us us to investigate the utility of employing alternate smearing schemes \citep{marzari1996ab}. Our experience suggests that while Gaussian and Methfessel-Paxton smearing \citep{methfessel1989high, Kresse_metal_semiconductor} schemes do indeed allow $N_x$ and $N_t$ to be lowered significantly (without otherwise impacting SCF convergence or changing energies and forces beyond acceptable tolerances), the parallel implementation of the CS2CF strategy sometimes suffers scalability and performance issues in these cases (due to a suboptimal number-of-cores to problem-size ratio). We therefore retain Fermi-Dirac smearing in the  calculations that follow.
}

We also remark that while the above results demonstrate the computational advantages of the CS2CF strategy for large systems, the strategy works equally well (i.e., in terms of lower computational wall times) for smaller systems. For example, for a $\textrm{SiDiamond3D}_{1 \times 1 \times 4}$ system ($32$ atoms, $64$ electrons) we found that the CS2CF strategy was able to reduce the (combined) subspace diagonalization and subspace rotation wall time to less than $0.001$ s, from $\sim 0.003$ s. The dense linear algebra operations for both strategies were carried out serially in this particular case. Hence, the CS2CF strategy appears to be a computationally advantageous replacement for the standard CheFSI strategy for a wide range of system sizes commonly encountered in Kohn-Sham calculations, as well as for sizes much larger.

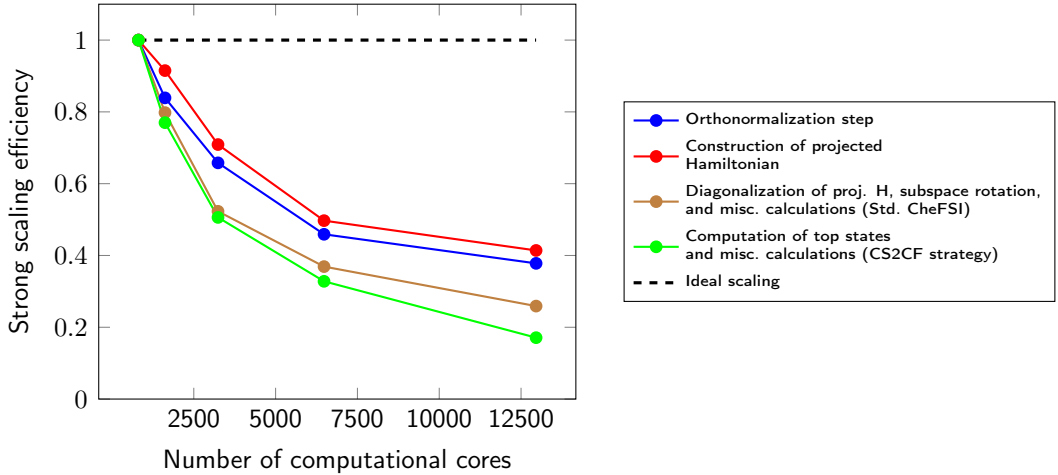
\begin{figure}
\subfloat{\scalebox{1.0}
{\begin{tikzpicture} 
\begin{axis}[
width=0.48\textwidth,
xlabel={Number of computational cores},
ylabel={Strong scaling efficiency},
legend style={at={(1.1,0.5)},anchor=west, font=\sffamily\tiny,fill=none,cells={anchor=west, align=left},row sep=0pt},
xtick={2500, 5000, 7500, 10000, 12500},
xticklabels={2500, 5000, 7500, 10000, 12500},
scaled x ticks = false,
label style={font=\sffamily\footnotesize},
tick label style={font=\sffamily\footnotesize},
ymin=0,ymax=1.1,
]
\addplot[color=blue,mark=*, thick] coordinates {
		(810, 1.00)
		(1620, 0.839)
		(3240, 0.658)
		(6480, 0.459)
		(12960, 0.378)
	};
%\addlegendentry{Orthonormalization step};	
\addplot[color=red,mark=*, thick] coordinates {
		(810, 1.00)
		(1620, 0.915)
		(3240, 0.709)
		(6480, 0.497)
		(12960, 0.414)
	};	
\addplot[color=brown,mark=*, thick] coordinates {
		(810, 1.00)
		(1620, 0.798)
		(3240, 0.523)
		(6480, 0.369)
		(12960, 0.259)
	};			
\addplot[color=green,mark=*, thick] coordinates {
		(810, 1.00)
		(1620, 0.77)
		(3240, 0.506)
		(6480, 0.328)
		(12960, 0.171)
	};		
\addplot[color=black, dashed, very thick] coordinates {
		(810, 1.00)
		(1620, 1.00)
		(3240, 1.00)
		(6480, 1.00)
		(12960, 1.00)
	};
\addlegendentry{Orthonormalization step};		
\addlegendentry{Construction of projected \\Hamiltonian};		
\addlegendentry{Diagonalization of proj. H, subspace rotation, \\and misc.~calculations (Std.~CheFSI)};		
\addlegendentry{Computation of top states \\and misc.~calculations (CS2CF strategy)};
\addlegendentry{Ideal scaling};
%\legend{Orthonormalization step, Construction of projected Hamiltonian, Rayleigh-Ritz $,$ Subspace Rotation and Misc.~calculations (Std.~CheFSI), Computation of top states and misc.~calculations (CS2CF strategy) };
%\legend{Orthonormalization step, Construction of projected Hamiltonian, Rayleigh-Ritz , Computation of top states  };
\end{axis}
\end{tikzpicture}}
}
\caption{\footnotesize{Strong scaling efficiencies of key steps in the
standard CheFSI and new CSF2CF strategies for the $\textrm{LiBCC3D}_{12
\times 12 \times 12}$ system. While the steps associated with CS2CF
scale less well than those associated with standard CheFSI, they yield a
significantly lower wall time (by a factor of $5.4$ for the case of
$12,960$ processors shown above). The steps common to both strategies
scale somewhat better, reaching approximately $40\%$ strong scaling
efficiency at $12,960$ processors in this case.  {The
parallel scaling property of the polynomial filter application step associated with $H$ (which appears in both standard CheFSI and CS2CF) is identical for both strategies and is not shown here (see previous work \citep{banerjee2016chebyshev} for details).}}}
\label{fig:strong_scaling}
\end{figure}

Next, we examine the strong parallel scaling properties of the CS2CF
strategy and contrast it with the standard CheFSI strategy. The parallel scaling properties of the polynomial filter application step associated with $H$ (that appears in both standard CheFSI and CS2CF) have been detailed in previous work\citep{banerjee2016chebyshev} and is identical for both strategies. Thus, we focus on the subspace problem construction and solution steps here. Taking the $\textrm{LiBCC3D}_{12 \times 12 \times 12}$ system as an example, we plot the strong scaling efficiency of the principal steps involved in constructing and solving the subspace problem via the standard CheFSI and CSF2CF strategies in Figure \ref{fig:strong_scaling}. We have used the data points corresponding to $810$ computational cores as the reference in this plot. 

From the plot, we see that the strong scaling efficiency of the steps
strictly associated with the CS2CF strategy is somewhat worse 
than that of the steps associated with the standard CheFSI
strategy, though not markedly so. This is attributed to the fact that the sizes of the matrices in the parallel dense linear algebra operations in the CS2CF approach are significantly smaller than those arising in the standard CheFSI approach. Thus, the PBLAS/ScaLAPACK routines do not parallelize as efficiently for the case of the CS2CF strategy. However, the steps associated with the CS2CF strategy execute significantly faster than their standard
CheFSI counterparts --- even for the case of $12,960$ computational cores, the wall time associated with the CS2CF strategy turns out to be lower than that of the standard CheFSI strategy by a factor of $5.4$. 
Thus, we may conclude that it is preferable
to use the CS2CF strategy regardless of the number of computational
cores allocated to the subspace problem.

It is also worth noting that the orthonormalization and projected
Hamiltonian construction steps common to both strategies fare somewhat better in terms of strong scaling performance (reaching approximately $40\%$ strong scaling efficiency at $12,960$ computational cores in this case). This suggests that it is worthwhile to allocate more computational cores to these parts of the calculation since the wall time for these steps can be reduced significantly.  We employ this strategy in the large benchmark calculations presented in the next section.

\subsection{Benchmark calculations on large systems} 
\label{sec:large_calcs}
From the results presented above, it is clear that 
the CS2CF strategy is well suited for
bringing down the computational wall times of large-scale Kohn-Sham 
calculations. We have already demonstrated
\citep{banerjee2016chebyshev} the superior computational efficiency of
the standard CheFSI strategy within DGDFT compared to existing
alternatives based on direct diagonalization (using ScaLAPACK) and 
certain sparse-direct solution strategies (namely,
PEXSI~\cite{CMS2009,JPCM_25_295501_2013_PEXSI,lin2014siesta}). 
Since the CS2CF strategy is successful in bringing down the wall times of the
standard CheFSI approach, it is the method of choice 
for large-scale calculations in DGDFT.
To demonstrate this, we display in Table \ref{tab:scf_wall_times} the wall time per SCF iteration for the large materials systems considered above, when the CS2CF strategy is employed.\footnote{In order to reduce the wall time associated with the subspace problem, we employed a larger number of computational cores than used in Figures~\ref{fig:Block_Chart_1} and \ref{fig:Block_Chart_2}. Also, rectangular process grid geometries (instead of nearly square ones) were used for carrying out parallel dense linear algebra operations.} {For comparison, we also show the corresponding wall times associated with direct diagonalization of the DG Hamiltonian using ELPA\citep{marek2014elpa, auckenthaler2011parallel} --- a state-of-the-art  massively parallel eigensolver library.}\footnote{{The interface to the ELPA library in DGDFT is provided through the ELectronic Structure Infrastructure (ELSI) framework.\citep{yu2018elsi}}} {The ELPA eigensolver was made to use the same total number of computational cores as the CS2CF strategy for all cases considered.}
\begin{table}
\begin{center}
{\normalsize{
\resizebox{\columnwidth}{!}{
\begin{tabular}{| c | c | c | c | c | c | c | c | c |}
\hline
& & & & & & & &\\
\multirow{3}{*}{System} & No.~of atoms & Total computational cores & ALB & Hamiltonian & CS2CF strategy & Total SCF wall time & Direct diagonalization & Total SCF wall time \\ & (No.~of electrons) & (cores used in subspace prob.~) & generation & update & (subspace prob.~time)  & via CS2CF strategy & via ELPA & via ELPA\\&  & [s] & [s] & [s] & [s] & [s] & [s] & [s]\\& & & & & & & &\\\hline
& & & & & & & &\\
{\large{$\textrm{Electrolyte3D}_{3 \times 3 \times 3}$}} & \large{$8,586$ ($29,808$)} & \large{$34,560$ ($3,456$)} & \large{$12$} & \large{$4$} & \large{$34$ ($19$)} & \large{$50$} & \large{$647$} & \large{$663$}\\
& & & & & & & &
\\\hline
& & & & & & & &\\
\large{$\textrm{SiDiamond3D}_{10 \times 10 \times 10}$} & \large{$8,000$ ($32,000$)} & \large{$34,560$ ($3,456$)} & \large{$9$} & \large{$2$} & \large{$40$ ($24$)} & \large{$51$} & \large{$648$} & \large{$659$}\\
& & & & & & & &
\\\hline
& & & & & & & &\\
\large{$\textrm{Graphene2D}_{8 \times 8}$} & \large{$11,520$ ($23,040$)} & \large{$27,648$ ($4,608$)} & \large{$4$} & \large{$2$} & \large{$35$ ($27$)} & \large{$41$} & \large{$262$} & \large{$268$}\\
& & & & &  & & &
\\\hline
& & & & & & & &\\
\large{$\textrm{CuFCC3D}_{10 \times 10 \times 10}$} & \large{$4,000$ ($44,000$)} & \large{$30,000$ ($3,000$)}  & \large{$20$} & \large{$9$} & \large{$75$ ($46$)} & \large{$104$} & \large{$199$} & \large{$228$}\\
& & & & & & & &
\\\hline
& & & & & & & &\\
\large{$\textrm{LiBCC3D}_{12 \times 12 \times 12}$} & \large{$27,648$ ($82,944$)} & \large{$38,880$ ($12,960$)} & \large{$22$} & \large{$13$} & \large{$180$ ($165$)} & \large{$215$} & \large{$5844$} & \large{$5879$}\\
& & & & & & & &\\\hline
\end{tabular}
}
}}
\end{center}
\caption{\footnotesize{SCF iteration wall times (in seconds, rounded to the nearest whole number) for large systems using the CS2CF strategy in DGDFT. The contributions of key computational steps are also shown. In the third and sixth columns, numbers in parenthesis indicate the numbers of processors used for subspace problem construction and solution, and wall times for those operations, respectively. {For comparison, corresponding wall times associated with direct diagonalization of the DG Hamiltonian using the ELPA library\citep{marek2014elpa, auckenthaler2011parallel} are also shown in the last two columns. ELPA was made to use the same total number of computational cores as the CS2CF strategy.}}}
\label{tab:scf_wall_times}
\end{table}

From the table we see that, with the ability to leverage large-scale computational resources, the CS2CF strategy within DGDFT is able to tackle several bulk and nano systems containing tens of thousands of electrons in less than a minute of wall time per SCF step. In terms of the number of electrons in the system, the $\textrm{LiBCC3D}_{12 \times 12 \times 12}$ case is the largest, and even in this case, the DGDFT-CS2CF methodology is able to complete each SCF iteration in a little over $3.5$ minutes of wall time. {In contrast, the wall times required for direct diagonalization of the DG Hamiltonian for these systems (via ELPA) are significantly longer, with the reductions achieved by the CS2CF strategy exceeding a factor of $15$ in some cases. The direct diagonalization wall time appears to be the closest to that of the CS2CF strategy for the $\textrm{CuFCC3D}_{10 \times 10 \times 10}$ system (likely because the overall size of the DG Hamiltonian is  smallest in this case); however, even in this situation, the CS2CF strategy appears to be faster by a factor of $\sim 2.6$. A comparison of the total SCF wall times between the CS2CF and direct diagonalization strategies also highlights significant gains, with an overall reduction factor of $\sim 13$ or higher in some cases.}

We reiterate that the wall times presented above
pertain to discretization parameter choices within DGDFT that lead to
well converged (chemically accurate) energies and forces. In our opinion, this is one of the key differences with earlier attempts at simulating such large scale metallic or semiconducting systems from first principles.
To further highlight this point, we employed the DGDFT-CS2CF methodology to carry out a $1.0$-ps \textit{ab initio} molecular dynamics simulation of the $\textrm{SiDiamond3D}_{10 \times 10 \times 10}$ system initialized at $300$ K ionic temperature. We used the NVE ensemble and a time step of $2.5$~fs for integrating the equations of motion using the velocity-Verlet scheme \citep{Hutter_abinitio_MD, Tadmor_Miller}. We initialized the system by randomly perturbing the positions of the silicon atoms in the $\textrm{SiDiamond3D}_{10 \times 10 \times 10}$ configuration and assigning the atoms random velocities consistent with the initial temperature. 
We then let the system evolve and equilibrate for $50$~fs before collecting data for $400$ ionic time steps (i.e., $1$~ps). 
To accelerate SCF convergence at each ionic step, we employed linear extrapolation of the real-space electron density and SCF converged wavefunctions from one ionic step were used as the starting point for  the SCF iterations on the next ionic step. Results from the simulation are shown in 
Fig.~\ref{fig:MD_Results}. The mean and standard deviation of the total energy (i.e., kinetic energy of the ions $+$ potential energy) are $-3.96329$ Ha per atom and $4.8\times10^{-6}$ Ha per atom, respectively. Additionally, the drift in total energy  (obtained using a linear fit) is less than $10^{-5}$ Ha/atom-ps. Thus, the scheme consistently produces high quality atomic forces and, consequently, excellent energy conservation.\footnote{{
Although atoms in the system were randomly perturbed before the start of the simulation, the particularly small drift may be due in part to the fact that the system is a crystal in the near harmonic regime. In other ab initio molecular dynamics simulations using the CS2CF strategy (involving the Electrolyte3D system, for example) we have found larger drifts for typical discretizations. In any case, since, as shown in Table~\ref{tab:scf_accuracy}, both energy and force errors associated with the CS2CF strategy are well below discretization error, drift for a given MD step size is controlled by discretization rather than eigensolution strategy.}} The mean ionic temperature during course of the simulation\footnote{During the course of molecular dynamics simulations using the NVE ensemble, it is usual for the system to reach a mean temperature somewhat different from the initial temperature\citep{Tadmor_Miller}.} comes out to be about $274$ K. 
\newsavebox{\mybox}
\savebox{\mybox}{
\tikzset{external/optimize=false}
  \begin{tikzpicture}[scale=0.45]
  \begin{axis}[
%width=0.48\textwidth,
y tick label style={
        /pgf/number format/.cd,
            fixed,
            fixed zerofill,
            precision=5,
        /tikz/.cd
    },
xlabel={Time (fs)},
ylabel={Total energy per atom (Ha)},
y label style={yshift=-5pt},
x label style={xshift=-10pt},
label style={font=\sffamily\footnotesize},
tick label style={font=\sffamily\footnotesize},
xmin=200,xmax=350,
ymin=-3.96338,ymax=-3.96322,
]
\addplot[thick, blue ] table[x index=0,y index=1]{./MD_Econ_Data.txt};
%\addplot[thick,mark=,blue] table[x index=0,y index=#3]{\datfile{#2}};
%\legend{Standard CheFSI (reference), CS Strategy};
\end{axis}  
  \end{tikzpicture}
}

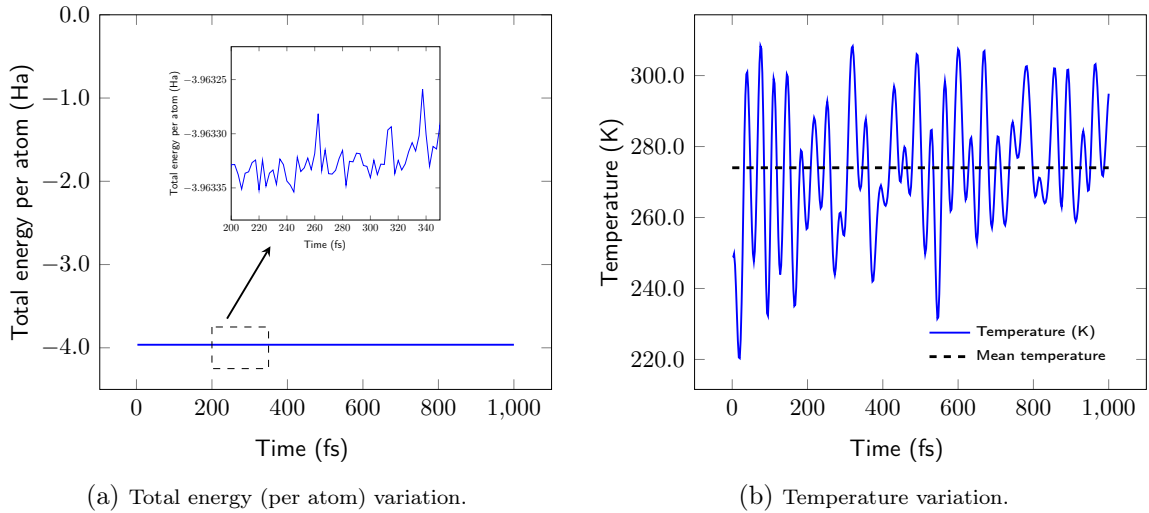
\begin{figure}
\subfloat[\scriptsize{Total energy (per atom) variation.} \label{subfig:MD_Econ}]{\scalebox{0.9}
{\begin{tikzpicture} 
%\pgfkeys{
%    /pgf/number format/precision=1, 
%    /pgf/number format/fixed zerofill=true
%}
\begin{axis}[
width=0.5\textwidth,
xlabel={Time (fs)},
ylabel={Total energy per atom (Ha)},
y label style={yshift=-5pt, font=\sffamily},
x label style={xshift=-10pt, font=\sffamily},
y tick label style={
        /pgf/number format/.cd,
            fixed,
            fixed zerofill,
            precision=1,
        /tikz/.cd
    },
label style={font=\sffamily\footnotesize},
tick label style={font=\sffamily\footnotesize},
ymin=-4.5,ymax=0,
]
\addplot[thick, blue ] table[x index=0,y index=1]{./MD_Econ_Data.txt};
\node[anchor=south west] at (1,-3) {\usebox{\mybox}};
%\draw [black, thick] (axis cs:-4.2,2) rectangle (axis cs:-3.8,3);
\draw [dashed, black]  (200,-3.75) rectangle (350,-4.25);
\draw [arrow, black ] (240,-3.65) -- (355,-2.78);
%\addplot[thick,mark=,blue] table[x index=0,y index=#3]{\datfile{#2}};
%\legend{Standard CheFSI (reference), CS Strategy};
\end{axis}
\end{tikzpicture}} 
}$\;\;$
\subfloat[\scriptsize{Temperature variation.}\label{subfig:MD_Temperature}]{\scalebox{0.9}
{\begin{tikzpicture} 
%\pgfkeys{
%    /pgf/number format/precision=1, 
%    /pgf/number format/fixed zerofill=true
%}
\begin{axis}[
width=0.5\textwidth,
xlabel={Time (fs)},
ylabel={Temperature (K)},
legend pos = {south west}, 
legend style={at={(0.5,0.12)},anchor=west, font=\sffamily\tiny,draw=none,fill=none,cells={anchor=west},row sep=0pt},
y label style={yshift=-5pt},
x label style={xshift=-10pt},
label style={font=\sffamily\footnotesize},
tick label style={font=\sffamily\footnotesize},
y tick label style={
        /pgf/number format/.cd,
            fixed,
            fixed zerofill,
            precision=1,
        /tikz/.cd
    },
]
\addplot[ blue, thick ] table[x index=0,y index=1]{./MD_Temperature_Data.txt};
\addplot[ black, very thick, dashed] coordinates {(0, 274.02) (1000, 274.02)};
\legend{Temperature (K), Mean temperature};
\end{axis}
\end{tikzpicture}}
}
\caption{\footnotesize{Results from $1.0$-ps NVE \textit{ab initio} molecular dynamics simulation of the $\textrm{SiDiamond3D}_{10 \times 10 \times 10}$ system ($8000$ atoms, $16,000$ electrons) using the CS2CF strategy in DGDFT. Total energy is well conserved, with a drift of less than $10^{-5}$ Ha/atom over the course of the simulation.}}
\label{fig:MD_Results}
\end{figure}

With the aid of the CS2CF strategy, the $1.0$-ps \textit{ab initio} MD simulation of the above 8,000-atom system can be carried out in $\sim$4.2 minutes per MD step, for a total of $\sim$28 hours of wall time on 34,560 computational cores. From the earlier discussion in Section~\ref{subsec:efficiency_and_scaling}, it is clear that doing a similar simulation without the use of the CS2CF strategy would have been far more expensive computationally, if not infeasible due to resource constraints. 
\section{Conclusions}
\label{sec:Conclusions}
In summary, we have presented a novel iterative strategy for 
KS-DFT calculations aimed at large system sizes, 
applicable to metals and insulators alike.
Our CS2CF methodology combines 
a complementary subspace (CS) strategy 
and two-level Chebyshev polynomial filtering (2CF) scheme 
to reduce subspace diagonalization to just fractionally occupied states and 
obtain those states in an efficient and scalable way, 
exploiting the limited spectral width and extremal nature of the resulting eigenvalue problem.
In so doing, the CS2CF approach reduces or eliminates 
some of the most computationally intensive and poorly scaling steps in 
large-scale Kohn-Sham calculations employing iterative solution methods.
We showed that the approach 
integrates well within the framework of 
the massively parallel discontinuous Galerkin electronic structure method. 
{Considering a variety of large systems, 
including metals, semimetals, semiconductors, and insulators, 
we then demonstrated that the use of
the CS2CF strategy within the massively parallel DGDFT code allows us
to tackle systems containing tens of thousands of electrons within a few
minutes of wall time per SCF iteration on large-scale computing
platforms.} {We found that the CS2CF strategy significantly 
outperforms alternatives based on direct diagonalization of the
Hamiltonian.} %, and we deem it to be the method of choice for large-scale calculations in DGDFT.} 
In particular, the strategy makes possible \textit{ab initio} molecular dynamics simulations of 
complex systems containing many thousands of atoms 
within a few minutes per MD step, as we demonstrate for bulk silicon. 

With the use of the CS2CF strategy, the
subspace diagonalization and subspace rotation steps cease to be the 
dominant parts of the calculation.
For the largest systems considered here, the time for carrying out
orthonormalization and forming the projected Hamiltonian then start
to contribute significantly. 
We aim to confront these next. 
Once again, our focus will not necessarily be on
lowering the computational complexity of these steps. 
Rather, any procedure that can lower the prefactor and/or 
improve the parallel scalability 
of these steps is likely be more effective in bringing down
wall times in practice, without sacrificing accuracy, 
thus pushing the envelope of 
\textit{ab initio} calculations 
to larger, more complex, and more realistic systems than feasible today. 
\begin{acknowledgement}
This work was performed, in part, under the auspices of the
U.S.~Department of Energy by Lawrence Livermore National Laboratory
under Contract DE-AC52-07NA27344 (J.E.P.). Support for this work was provided through Scientific Discovery through Advanced Computing (SciDAC) program funded by U.S.~Department of Energy, Office of Science, Advanced Scientific Computing Research and Basic Energy Sciences (A.S.B., L.L., C.Y., and J.E.P), by the Center for Applied Mathematics for Energy Research Applications
(CAMERA), which is a partnership between Basic Energy Sciences and
Advanced Scientific Computing Research at the U.S Department of Energy
(L. L. and C. Y.), and by the National Science Foundation
under Grant No. 1450372 and the Department of Energy under Grant No.
DE-SC0017867 (L. L.). We would like to thank the National Energy Research Scientific Computing (NERSC) center for making computational resources available to them. A.S.B. would like to thank Meiyue Shao, Roel Van Beeumen and Wei Hu (Lawrence Berkeley Lab) for informative discussions and for their help with improving the presentation of the manuscript. We thank Mitchell Ong for providing the lithium-ion electrolyte system and Donald Hamann for his assistance in generating soft and robust ONCV potentials. {Finally, we would like to thank the anonymous reviewers for their comments and suggestions, which helped to improve the manuscript at a number of points.}
\end{acknowledgement}

\footnotesize{
\bibliography{Complementary_Subspace_Strategy}
}

\end{document}